\documentstyle[11pt,epsf]{article}
\def\EllipticE{{\bf E}}         % standard elliptic integrals
\def\EllipticK{{\bf K}}         % E and K
\def\0{{\emptyset}}             % symbol for vacancy
\def\inp{p}                     % symbol for input rate
\def\conc{c}                    % local particle concentration discrete
\def\dens{n}                    % particle density mean field
\def\pa{r}                      % first  bulk-rates parameter
\def\pb{s}                      % second bulk-rates parameter
\def\pc{t}                      % third  bulk-rates parameter
\def\pac{\hat{r}}               % first  bulk-rates parameter in continuous
\def\pbc{\hat{s}}               % second bulk-rates parameter in continuous
\def\density{\rho}              % particle density continuum limit
\def\disp{\mu}                  % dispersion function
\def\spacing{\lambda}           % lattice spacing
\def\be{\begin{equation}}
\def\ee{\end{equation}}
\newcommand{\ba}{\begin{eqnarray}}
\newcommand{\ea}{\end {eqnarray}}

\renewcommand{\theequation}{\arabic{section}.\arabic{equation}}
\begin{document}
\pagestyle{headings}
%----------------------------------------------------------------------
\setcounter{page}{1}
\pagestyle{plain}
\setcounter{equation}{0}
%
%
%----------------------------------------------------------------------
% Title page:
%----------------------------------------------------------------------
%
\ \\[12mm]
\begin{center}
     {\large \bf UNIVERSALITY PROPERTIES OF THE STATIONARY STATES \\[2mm]
                 IN THE ONE-DIMENSIONAL COAGULATION-DIFFUSION MODEL \\[2mm]
                 WITH EXTERNAL PARTICLE INPUT}
	     \\[20mm]
\end{center}
\begin{center}
\normalsize
	Haye Hinrichsen$^{\diamond}
	\footnote{e-mail: \tt fehaye@wicc.weizmann.ac.il}$, 
	Vladimir Rittenberg$^{\star},
	\footnote{e-mail: \tt unp01C@ibm.rhrz.uni-bonn.de}$ 
	and Horatiu Simon$^{\star}
	\footnote{e-mail:  \tt hsimon@theoa1.physik.uni-bonn.de}$ \\[13mm]
	$^{\diamond}$ {\it Department of Physics of Complex Systems\\
	Weizmann Institute of Science\\ Rehovot 76100, Israel}\\[4mm]
	$^{\star}$ {\it Universit\"{a}t Bonn,
	Physikalisches Institut \\ Nu\ss allee 12,
   D-53115 Bonn, Germany}
\end{center}
\vspace{15mm}
{\bf Abstract:}
We investigate with the help of analytical and numerical methods
the reaction $A+A \rightarrow A$ on a one-dimensional lattice opened 
at one end and with an input of particles at the other end. 
We show that if the diffusion rates to the left and to the right are equal,
for large $x$, the particle concentration $\conc(x)$ behaves like $A_s x^{-1}$
($x$ measures the distance to the input end). 
If the diffusion rate in the direction pointing away from the source is larger 
than the one corresponding to the opposite direction the particle concentration
behaves like $A_a x^{-1/2}$. 
The constants $A_s$ and $A_a$ are independent of the input and the two 
coagulation rates. The universality of $A_a$ comes as a surprise since in the 
asymmetric case the system has a massive spectrum.
\\[15mm]
cond-mat/9606088\\[2mm]
\rule{6.6cm}{0.2mm}
\begin{flushleft}
\parbox[t]{3.5cm}{\bf Key words:}
\parbox[t]{12.5cm}{Non-equilibrium statistical mechanics, 
                   reaction-diffusion systems,\\ coagulation model,
                   universality}
\\[2mm]
\parbox[t]{3.5cm}{\bf PACS numbers:}
\parbox[t]{12.5cm}{05.40.+j, 05.70.Ln}
\end{flushleft}
\normalsize
\thispagestyle{empty}
\mbox{}
\pagestyle{plain}
%
%----------------------------------------------------------------------
% 1) INTRODUCTION:
%----------------------------------------------------------------------
%
\newpage
\setcounter{page}{1}
\setcounter{equation}{0}
\section{Introduction}
\label{Intro}
The study of one-dimensional reaction-diffusion models far from 
thermal equilibrium is a field of growing interest. 
The dynamics of these models is characterized by non-trivial
correlations so that ordinary mean field techniques fail. 
Therefore theoretical descriptions have to take local fluctuations 
into account. In general this is a very difficult task and
approximation techniques are needed. However, there is a small 
number of {\it exactly solvable} models where we can derive
exact results. The great interest in solvable models comes
from the fact that their physical properties 
appear also in many other more complicated models. 
If we consider one-dimensional two-state models (i.e. 
models with only one species of particles), there are only
two classes of exactly solvable systems. 
The first class includes diffusion (or exclusion) models which are 
solvable by means of Bethe ansatz techniques. 
The second class contains mainly models with an 
underlying theory of free fermions. The most important representative
of this class is the so-called coagulation model
which is the subject of the present work.

Coagulation models describe particles which diffuse
stochastically in a $d$-dimensional space. 
When two particles meet at the same place they coalesce to a single
one $(A+A \rightarrow A)$. 
For a possible experimental realization see
\cite{Experiment}. The theoretical study of coagulation models has
a long history. It started with the observation that the critical
dimension of the corresponding field theory is $d_c=2$ \cite{Fluctuations}.
A breakthrough towards the exact solution of the one-dimensional
coagulation model on a lattice was the introduction 
of so-called interparticle distribution
functions (IPDF's)  \cite{IPDF}. 
For certain reaction-diffusion processes the IPDF formalism leads to a 
hierarchy of decoupled differential equations similar to those obtained
for the Glauber model \cite{Glauber}. The most general conditions for the
decoupling to occur as well as the cases when the underlying Hamiltonian
can be diagonalized in terms of free fermions are given in Ref. \cite{PRS}.
A variety of exact solutions
were found for the coagulation model with or without back reaction
(decoagulation $A\rightarrow A+A$) \cite{IPDF}.

\noindent
The one-dimensional coagulation model with 
spatial {\it homogeneous} external particle 
input at all sites has been studied extensively \cite{GlobalInput} and
algebraic relaxation times have been observed.
In this paper we investigate the same model with open boundary 
conditions and {\it localized} particle input at the ends of the chain.
This problem was considered for the first time in Ref. \cite{PointLikeSource}
where various scaling relations could be obtained. In the present work
we compute the particle concentration in the stationary state.
We give the full solution on the lattice and in the continuum. 
The main motivation of this paper stems from the observation that in the
mean field approximation (to be reviewed later) the density of particles has 
an algebraic decay and one can look if one has universality properties. 

The coagulation model studied in this paper is defined as follows.
Particles of one species diffuse stochastically on a linear 
one-dimensional  lattice with $L$ sites. The diffusion may be biased due to
some external force.
If two particles meet at the same site, they coalesce to a 
single one. In addition particles are added stochastically 
with a given probability 
at the endpoints of the lattice. We use random sequential updates, 
i.e. we assume continuous time evolution which is described
by a linear master equation. Altogether the dynamics is defined
by the six nearest neighbor processes with rates shown in Figure \ref{f0}.

%
%************************************
% Pictures describing the processes:
%************************************
%
\def\latt{\thinlines\put(0,1){\line(1,0){9}}
			 \put(0,6){\line(1,0){9}}\thicklines}
\def\lattl{\thinlines\put(2,1){\line(1,0){7}}
			 \put(2,6){\line(1,0){7}}\thicklines}
\def\lattr{\thinlines\put(0,1){\line(1,0){7}}
			 \put(0,6){\line(1,0){7}}\thicklines}
\def\arr{\put(3,5){\vector(1,-1){3}}}
\def\arl{\put(6,5){\vector(-1,-1){3}}}
\def\dol{\put(2,5){\vector(0,-1){3}}}
\def\dor{\put(7,5){\vector(0,-1){3}}}
\def\occ{\circle*{1}}
\def\vac{\circle{1}}
\def\picar{\begin{picture}(18,7)\latt \arr
\put(2,6){\occ} \put(7,6){\vac} \put(10,3){$a_R$}
\put(2,1){\vac} \put(7,1){\occ} \end{picture}}
\def\pical{\begin{picture}(18,7)\latt \arl
\put(2,6){\vac} \put(7,6){\occ} \put(10,3){$a_L$}
\put(2,1){\occ} \put(7,1){\vac} \end{picture}}
\def\piccr{\begin{picture}(18,9)\latt \arr \dor
\put(2,6){\occ} \put(7,6){\occ} \put(10,3){$c_R$}
\put(2,1){\vac} \put(7,1){\occ} \end{picture}}
\def\piccl{\begin{picture}(18,9)\latt \arl \dol
\put(2,6){\occ} \put(7,6){\occ} \put(10,3){$c_L$}
\put(2,1){\occ} \put(7,1){\vac} \end{picture}}
\def\picinpr{\begin{picture}(18,9)\lattr \dor
\put(7,6){\vac} \put(10,3){$\inp_R$}
\put(7,1){\occ} \end{picture}}
\def\picinpl{\begin{picture}(18,9)\lattl \dol
\put(2,6){\vac} \put(10,3){$\inp_L$}
\put(2,1){\occ} \end{picture}}
\def\pichopp{\begin{picture}(18,7)\put(1,3)
{biased diffusion}\end{picture}}
\def\piccoag{\begin{picture}(18,7)\put(1,3)
{biased coagulation}\end{picture}}
\def\picinput{\begin{picture}(18,7)\put(1,3)
{particle input at the boundaries}\end{picture}}

\begin{figure}
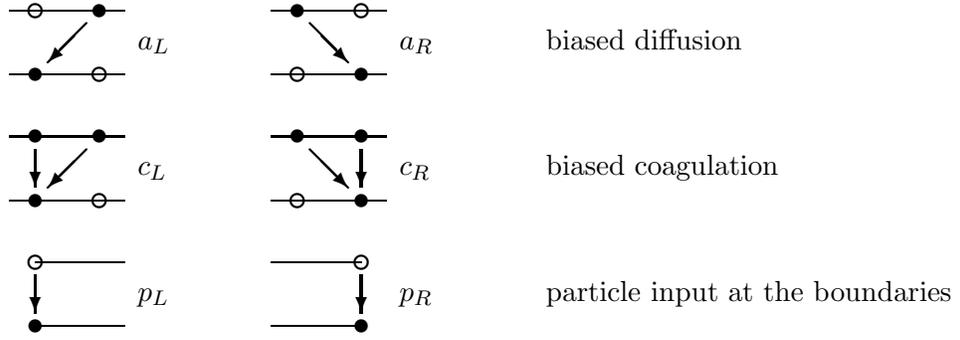

\begin{center}
\setlength{\unitlength}{1.7mm}
\begin{tabular}{ccl}
\pical & \picar & \pichopp \\
\piccl & \piccr & \piccoag \\
\picinpl & \picinpr & \picinput \\
\end{tabular}
\end{center}
\caption{Bulk and boundary processes in the coagulation-diffusion model}
\end{figure}

\noindent
The master equation for a lattice of $L$ sites can be written in a compact 
form by (for notations c.f. Ref. \cite{Alcaraz})
\begin{equation}
\label{EqOfMotion}
\frac{\partial}{\partial t}\,|P(t)\rangle \;=\; -H |P(t)\rangle
\end{equation}
where the vector $|P(t)\rangle$ denotes the probability distribution.
$H$ is the time evolution operator which
can be written as a sum of nearest-neighbor reaction matrices
$H_{n,n+1}$ plus two further matrices for
particle input $I_L$ and $I_R$:
\begin{equation}
H \;=\; I_L \,+\, I_R \,+\, \sum_{n=1}^{L-1} \, H_{n,n+1}\,.
\end{equation}
\noindent
In the canonical basis of particle configurations
$(\,|\0\0\rangle,\,|\0A\rangle,\,|A\0\rangle,\,|AA\rangle)$
these matrices read
\begin{equation}
\label{Hamiltonian}
H_{n,n+1} \;=\; \left(\begin{array}{cccc}
		   \,\,\,0\,\, & 0    & 0    & 0    \\
		   0 & a_L  & -a_R & -c_R \\
		   0 & -a_L & a_R  & -c_L \\
		   0 & 0    & 0    & c_L+c_R
		  \end{array} \right)_{n,n+1}\,
\end{equation}
\noindent
and
\begin{equation}
I_L  \;=\; \left(\begin{array}{cc} \inp_L & 0 \\ -\inp_L & 0
		  \end{array} \right)_1
\hspace{30mm}
I_R  \;=\; \left(\begin{array}{cc} \inp_R & 0 \\ -\inp_R & 0
		  \end{array} \right)_L \,.
\label{hlhr}
\end{equation}
It is useful to introduce some notations
\begin{equation}
\pa \;=\; \frac{2\,(a_R-a_L)}{(a_R+a_L)}\,,
\hspace{5mm}
\pb \;=\; \frac{2\,(c_R+c_L)}{(a_R+a_L)}\,,
\hspace{5mm}
\pc \;=\; \frac{2\,(c_R-c_L)}{(a_R+a_L)}\,,
\hspace{5mm}
q \;=\; \Biggl(\frac{a_R}{a_L}\Biggr)^{1/2}
\label{paramdisc}
\end{equation}
\noindent
$c(i)$ denotes the particle concentration at the site $i$. When we will
consider the continuum limit (the lattice spacing $\spacing\,\rightarrow\,0$),
we will use the notation 
\begin{equation}
\density(x)\,=\,\spacing^{-1} c\Biggl(\frac i {\spacing}\Biggr)\,,
\hspace{5mm}
\pac \;=\; \frac{\pa}{\spacing}\,,
\hspace{5mm}
\pbc \;=\; \frac{\pb}{\spacing}\,,
\hspace{5mm}
\hat{p}\;=\;\frac{p_L}{\spacing^2}
\label{paramcont}
\end{equation}
Before giving our results we remind the reader of the improved  mean field 
calculations in Ref.\cite{PointLikeSource}. Assuming that the coagulation 
rate $\pbc$ is proportional to the concentration $\pbc\,=\,\zeta\density(x)$,
for large values of $x$ (we take $p_R=0$ and the source is at $x=0$), the
densities are
\begin{equation}
\begin{array}{lll}
q<1 & \mbox{(bias to the left) :} & \hspace{10mm}
\density(x) \propto e^{-|\pac |x} \\[1mm]
q=1 & \mbox{(symmetric diffusion) :} & \hspace{10mm}
\density(x) \approx \sqrt{\frac{2}{\zeta}}\,x^{-1} \\
q>1 & \mbox{(bias to the right) :} & \hspace{10mm}
\density(x) \approx \sqrt{\frac{\pac}{2\zeta}}\,x^{-1/2} 
\end{array}
\end{equation}
Notice the algebraic fall-off for $q\geq 1$.
\\
\noindent
The IPDF method is applicable if 
\be
\label{FC}
a_R=c_R=q\,, \hspace{20mm}
a_L=c_L=q^{-1}\,.
\ee
These conditions are equivalent to $\pa\,=\,\pc$ and $\pb\,=\,2$.
This will also be called the fermionic case (the Hamiltonian can be 
diagonalized in terms of free fermions \cite{PRS,Krebs,Peschel}). The special
case $q=1\;,p_R\,=\,p_L\,=\,\infty\,$ was already studied in a paper by 
Derrida et al \cite{PointlikeSourceExact} which was a source of inspiration for
the present work. We list now our main results in the $p_R=0$ case:
\\
\vspace{3mm}

\noindent
a) {\em Lattice in the thermodynamical limit} ($j$ fixed,$\;L\rightarrow
\infty$)
\begin{eqnarray}
\label{OOXD}
&&c(j) \;=\; \frac{2}{\pi j}+\frac{1}{\pi j^2} -\frac{1}{2 \pi j^4} 
+\Biggl(\frac{3}{8 \pi} - \frac{12}{\pi p_L^2}\Biggr)\frac 1 {j^5} + 
O(j^{-6})
%{\cal O}(j^{-6})
\hspace{45mm} q=1 \\[2mm]
\label{AP}
&&c(j) \;=\;  \sqrt{\frac{q^2-1}{(q^2+1)\pi j}}
\,\Biggl[ 1 + \Biggl(\frac{3q^4+20q^2-1}{8(q^4-1)} -
\frac{(q^2-1)^3}{2q^2(q^2+1)p_L^2 }\Biggr)\frac 1 j \,\Biggr] + 
O(j^{-5/2})
%{\cal O}(j^{-5/2})
\hspace{8mm} q>1 \hspace{12mm} \\[2mm]
&& c(j) \;=\;  q^{4j} \left(
\sqrt{\frac{1-q^2}{(q^2+1)\pi j}} + 
O(j^{-3/2}) \right)
%{cal O}(j^{-3/2}) \right)
\hspace{70mm} q<1
\end{eqnarray}

Several exact values for $c(j)$ are given in Appendix A.
\\
\vspace{3mm}

\noindent
b) {\em Continuum and thermodynamical limit} ($x$ fixed,$\;L\rightarrow
\infty$)
\begin{eqnarray}
\label{Concon}
&& \density (x) \;=\;
\frac 2  {\pi x}\,-\,\frac{12}{\pi \hat{p}^2 x^5} +
\frac{5040}{\pi \hat{p}^4 x^9} + O(\hat{p}^{-6}x^{-13})
\hspace{27mm} q=1
\\[2mm]
\label{Iqsf1}
&& \density (x)\;=\;\sqrt{\frac \pac  {2\pi x}} \,+\,\frac 1 {4\sqrt{2\pi \pac }}
\Biggl[ \frac {11} 2 \,-\,\frac{\pac ^4}{\hat{p}^2} \Biggr]
\frac 1 {x^{3/2}}\;+\;O(x^{-5/2})
\hspace{11mm} q>1
\end{eqnarray}
\vspace{3mm}

\noindent
c) {\em Continuum and scaling limit} ($z=\frac x L$ fixed,$\;L\rightarrow
\infty,\; q=1$)
\be
\lim_{\textstyle{L\rightarrow \infty \atop \textstyle {x\rightarrow \infty
\atop z=\frac x L \;{\mbox fixed}}}}
\,L\,\density(x) \;=\;\Phi(z)
\ee
\be
\Phi(z)\;=\;
\frac{\sinh(\pi z)\,+\,
\sin(\pi z)}{\cosh(\pi z)\,-\,\cos(\pi z)}\;+\;
\sum_{k=1}^{+\infty}\Biggl\{\frac{\sinh(2\pi(z/2-k))\,+\,
\sin(\pi z)}{\cosh(2\pi(z/2-k))\,-\,\cos(\pi z)}\;\;+\;(\;k\;
\longleftrightarrow \;-k\;)\Biggr\}
\label{SF1}
\ee
\vspace{1mm}

\noindent
As a by-product, the one-hole functions are also obtained. This result is not
trivial since they represent two-point correlation functions.

In the long Section \ref{ExactSolution} and in the 
Appendices A and B these results
are derived. Many of our calculations are extensions of results obtained in
references \cite{Krebs} and \cite{Peschel} from which we borrow the notations.
Also in Section \ref{ExactSolution} we give the spectrum of the Hamiltonian
in the one-hole sector. For $q\neq1$ the spectrum is massive in spite of the 
algebraic behavior seen in Eqs. (\ref{AP}) and (\ref{Iqsf1}). For $q=1$ most
of the excitations are massless (they coincide with those of the open chain
($p_L=0$)) but there are also some massive excitations with a mass given by 
$p_L$. In many systems time-like and space-like properties seem to be coupled
in the sense that long range correlations in time imply long range 
correlations in space. This is not necessary valid for stochastic models which
are not isotropic. As we will see in this model one can have short range 
correlations
in time but long range correlations in space (for another example, see Ref.
\cite{Alcaraz}).

In Section \ref{Numerical results} we consider the problem of the universality:
The coefficient $\frac 2 {\pi}\;$ for the leading contribution in the
$q=1$ case (see Eqs. (\ref{OOXD}) and (\ref{Concon})), the coefficient
$\sqrt{\frac{q^2-1}{(q^2+1)\pi}}\;$ for $q>1$ (see Eqs. (\ref{AP}) and
(\ref{Iqsf1})) and the finite-size scaling function (\ref{SF1}).
For this purpose we keep the definition $\sqrt{\frac{a_R}{a_L}}\,=\,q\;$
but leave the coagulation rates $c_R$ and $c_L$ arbitrary. For the open chain
(no input) the spectra are known to be massless for $q=1$ and massive for
$q>1$ \cite{Alcaraz}. The modifications introduced by the boundary terms
are supposed not to change radically the picture. Using Monte Carlo
simulations (the details are given in Appendix C) we show that indeed for
several values of $c_L$ and $c_R$ the expansion coefficients as well
as the finite-size scaling functions are universal. 

The reader not interested in lengthy calculations 
can skip Section \ref{ExactSolution} and proceed
directly to Section \ref{Numerical results}.
%
%
%
%----------------------------------------------------------------------
% 3) Exact solution
%----------------------------------------------------------------------
%
\section{Exact solution in the $a_R\,=\,c_R\,=\,q,\,a_L\,=\,c_L\,=\,q^{-1}$ 
case.}
\label{ExactSolution}
\subsection{Finite lattice calculations}
\label{Finitelatticecalculations}
\noindent
In this Section we give the full solution of the coagulation model
with particle input at the boundaries. We use the IPDF formalism
\cite{IPDF} in which the whole problem is
formulated in terms of probabilities for finding sequences of
unoccupied sites (holes). In this basis the master equation
leads to a hierarchy of sets of equations according the number of holes. 
It is known \cite{PRS} that these sets decouple from the higher ones provided 
that the rates for diffusion and coagulation coincide  (see Eq. (\ref{FC})).
Therefore the one-hole sector decouples from the higher sectors 
and can be solved separately. 
In what follows we will assume that the above condition holds.

For completeness we will consider the model with input at the left end (rate
$p_L$) and right end (rate $p_R$). \footnote{In order to avoid confusion in
terminology we give in parenthesis an alternative denomination used in the
literature: periodic boundary conditions (model on a ring), open boundary
conditions (linear chain with closed ends), open boundary with particle input
(chain with open ends).} We will actually show that by solving the
problem with $p_R\,=\,0$ one can obtain the general solution $p_R\,\neq\,0$.
Although in principle feasible, we didn't look to the case when one has also
output of particles at both end.

Although not obviously needed for the study of the stationary state, we will
also give the spectrum of the problem in the one-hole sector for two reasons.
One is technical: the eigenfunctions and eigenvalues occur in the expression
of the stationary state hole probabilities. The second one is related to the
physical significance of our result: it is important to know whether one has
massless or massive excitations.

Let $\Omega(j,m,t)$ denote the probability to find the sites
$j+1, j+2, \ldots m$ empty at time $t$. By a careful analysis
of the elementary processes taking place at the edges of the
hole one is led to the following equations of motion for the
one-hole sector:
\begin{itemize}
\item for holes which do not touch the boundaries
($0<j<m<L$):
\begin{eqnarray}
\label{FullEquations}
\frac{d}{dt}\Omega(j,m,t) &=&
q\,\Omega(j-1,m,t) \,+\,
q^{-1}\,\Omega(j+1,m,t) \\
&& +\,
q\,\Omega(j,m-1,t) \,+\,
q^{-1}\,\Omega(j,m+1,t)
\,-\, 2\,(q+q^{-1}) \,\Omega(j,m,t) \nonumber
\end{eqnarray}
\noindent
\item for holes touching the left boundary
($0=j<m<L$):
\begin{equation}
\label{LeftBoundaryCondition}
\frac{d}{dt}\Omega(0,m,t) \;=\;
q\,\Omega(0,m-1,t) \,+\,
q^{-1}\,\Omega(0,m+1,t) \,-\, \nonumber
(q+q^{-1}+\inp_L)\, \Omega(0,m,t)
\end{equation}
\noindent
\item for holes touching the right boundary
($0<j<m=L$):
\begin{equation}
\frac{d}{dt}\Omega(j,L,t) \;=\;
q \Omega(j-1,L,t) +
q^{-1} \Omega(j+1,L,t) \,-\, \nonumber
(q+q^{-1}+\inp_R)\, \Omega(j,L,t)
\end{equation}
\noindent
\item for the hole extending over the whole chain
($j=0, m=L$):
\begin{equation}
\label{FullEquationsEnd}
\frac{d}{dt}\Omega(0,L,t) \;=\;
-(\inp_L+\inp_R) \,\Omega(0,L,t)\,.
\end{equation}
\noindent
\end{itemize}
In these equations we have taken $\Omega(j,j,t)=1$.
This leads to an {\it inhomogeneous}
system of equations. Separating the time dependence and
introducing rescaled probabilities 
\begin{equation}
\label{Rescaling}
\Omega(j,m,t) \;=\; e^{-\Lambda t} q^{+j+m} \, \tilde{\Omega}(j,m) 
\end{equation}
\noindent
we obtain the simplified system of equations
\begin{eqnarray}
\label{ScaledBulk}
\Bigl(2(q+q^{-1})-\Lambda \Bigr) \tilde{\Omega}(j,m)
&=& \tilde{\Omega}(j-1,m)+\tilde{\Omega}(j+1,m)
+\tilde{\Omega}(j,m-1)+\tilde{\Omega}(j,m+1)
\hspace{10mm}
\\[1mm]
\label{ScaledLeft}
\Bigl(q+q^{-1}+\inp_L-\Lambda\Bigr)\, \tilde{\Omega}(0,m)
&=& \tilde{\Omega}(0,m-1)+\tilde{\Omega}(0,m+1)
\\[1mm]
\label{ScaledRight}
\Bigl(q+q^{-1}+\inp_R-\Lambda\Bigr) \,\tilde{\Omega}(j,L)
&=& \tilde{\Omega}(j+1,L)+\tilde{\Omega}(j-1,L)
\\[1mm]
\label{ScaledCorner}
\label{ScaledEquationsEnd}
\Bigl(\inp_L+\inp_R-\Lambda\Bigr)\, \tilde{\Omega}(0,L) &=& 0
\end{eqnarray}
\noindent
with the inhomogeneous boundary condition
$\tilde{\Omega}(j,j)\;=\; q^{-2j}$.
\\[3mm]
\indent
The {\it homogeneous} set of solutions describes the relaxational
modes of the system. It is obtained by setting $\tilde{\Omega}(j,j)=0$ 
and can be computed easily by using similar
techniques as in Ref. \cite{Krebs} which rely mainly on the
invariance of the bulk equation (\ref{FullEquations})
under reflections $j \leftrightarrow m$ and
$j \leftrightarrow L-m$. Denoting
\be
g(j,z) \;=\; \frac {\sinh \Bigl( j \,
	     \mbox{arcsinh} \frac{1}{2} (q+q^{-1}-z) \Bigr) }
	    {\sinh \Bigl( L \,
	     \mbox{arcsinh} \frac{1}{2} (q+q^{-1}-z) \Bigr) }
\ee
the homogeneous solutions are
\begin{eqnarray}
\label{PhiZeroMode}
\Phi_0(j,m) &=& g(L-j,p_L)\,g(m,p_R)-g(L-m,p_L)\,g(j,p_R)\\
\Phi^{(L)}_k(j,m) &=& \sin\frac{\pi k j}{L} \, g(L-m,p_L) -
\sin\frac{\pi k m}{L} \, g(L-j,p_L)  \\
\Phi^{(R)}_k(j,m) &=& \sin\frac{\pi k j}{L} \, g(m,p_R)-
\sin\frac{\pi k m}{L} \, g(j,p_R)\\
\label{PhiTwoFermions}
\Phi_{k,l}(j,m) &=& \sin\frac{\pi k j}{L}\,\sin\frac{\pi l m}{L}\,-\,
\sin\frac{\pi k m}{L}\,\sin\frac{\pi l j}{L}\,.
\end{eqnarray}
They have the excitation energies
\begin{eqnarray}
\label{Oferm}
\Lambda_0 & = & \inp_L+\inp_R\\[2mm]
\label{aferm}
\Lambda^{(L)}_k & = & q+q^{-1}+\inp_L -
2 \cos\frac{\pi k}{L}\\[2mm]
\label{bferm}
\Lambda^{(R)}_k &=& q+q^{-1}+\inp_R -
2 \cos\frac{\pi k}{L}\\[2mm]
\label{tferm}
\Lambda_{k,l} &=& 2(q+q^{-1})\,-\,2\,
\cos\frac{\pi k}{L}\,-\,2\,\cos\frac{\pi l}{L}
\end{eqnarray}
\noindent
where $1 \leq k < l \leq L$.  
In contrast to the coagulation model without particle input 
($p_R\,=\,p_L\,=\,0$) where the spectrum is massless for $q=1$ and massive
otherwise ($q$ real), in the case of particle input, the spectrum is 
more complex. Even for $q=1$ where most of the excitations are massless
(Eq.(\ref{tferm})) we get some massive ones too (Eqs.(\ref{Oferm}) - 
(\ref{bferm})).
\\[3mm]
\indent
The derivation of the {\it inhomogeneous} (steady state) solution
is more difficult. For symmetric coagulation
on a ring with infinite particle input at a single site
an exact solution has been found recently in Refs.
\cite{PointlikeSourceExact,Potts}. This solution applies
to an open chain with symmetric diffusion (q=1)
and infinite particle input at both ends
($\inp_L=\inp_R=\infty$). It is given by
$\tilde{\Omega}(j,j)=1$ and
\begin{equation}
\label{DerridaSolution}
\tilde{\Omega}(j,m) =
\frac{8}{L^2} {\sum_{k,l=1}^{L-1}}'\,
\frac{ \sin\frac{\pi k}{L}\sin\frac{\pi l}{L}\,
      (\sin\frac{\pi k j}{L}\sin\frac{\pi l m}{L} -
       \sin\frac{\pi k m}{L}\sin\frac{\pi l j}{L}) }
     { (\cos\frac{\pi l}{L}-\cos\frac{\pi k}{L})
       (2-\cos\frac{\pi k}{L}-\cos\frac{\pi l}{L}) }
\hspace{10mm}
(j<m)
\end{equation}
where the prime indicates that
the sum runs only over even values of $k$ and odd
values of $l$. Formally this solution can be expressed in terms
of the two-particle excitations (\ref{PhiTwoFermions}) by
\begin{equation}
\label{GeneralExpression}
\tilde{\Omega}(j,m) \;=\;
\frac{8}{L^2} \,
\sum_{k,l=1}^{L-1} \frac{f_{k,l}\,\Phi_{k,l}(j,m)}{\Lambda_{k,l}}
\end{equation}
where
\begin{equation}
\label{DerridaStructureFunction}
f_{k,l} \;=\;
\frac12\,(1-(-)^{k+l})\,
\frac{\sin\frac{\pi k}{L}\sin\frac{\pi l}{L}}
     {\cos\frac{\pi l}{L}-\cos\frac{\pi k}{L}}\,.
\end{equation}
plays the role of a structure function. However, we one can prove 
that the general stationary solution
for the asymmetric diffusion and finite particle input rates
has the same structure and differs only in the structure function $f_{k,l}$.
This function can be derived as follows. Let us symbolize
a contraction of two functions over momentum indices $k,l$ 
by $\langle \cdot,\cdot \rangle_{k,l}$ and similarly a contraction over
spatial indices by $\langle \cdot,\cdot \rangle_{j,m}$. 
Then Eq. (\ref{GeneralExpression}) reads
$\tilde{\Omega}=\langle f,\frac{\Phi}{\Lambda} \rangle_{k,l}$ 
and therefore the application of the discretized Laplacian 
$\Delta \Phi=\Lambda\Phi$ yields $\Delta \tilde{\Omega}=
\langle f,\Phi \rangle_{k,l}$ which is zero everywhere except 
at the boundaries. Using the orthogonality relation
$\langle \Phi_{k,l},\Phi_{k',l'} \rangle_{j,m} \sim \delta_{k,k'}\delta_{l,l'}$
one can therefore compute $f$ by
\begin{equation}
f \sim \langle f,\delta \delta\rangle_{k,l}
  \sim \langle f,\langle \Phi, \Phi \rangle_{j,m} \rangle_{k,l}
  \sim \langle \langle f, \Phi \rangle_{k,l} , \Phi \rangle_{j,m}
  \sim \langle \Delta \tilde{\Omega} , \Phi \rangle_{j,m} \,.
\end{equation}

\noindent
Carrying out these contractions it turns out that the structure
function $f$ consists of three parts
\begin{equation}
\label{ThreeParts}
f_{k,l} \;=\; f_{k,l}^{(\infty)} \,+\,f_{k,l}^{(L)} \,+\,f_{k,l}^{(R)}\,.
\end{equation}
The first part $f^{(\infty)}_{k,l}$ describes the asymmetric coagulation
model with infinite particle input rates at {\it both} ends. It is given by
\begin{equation}
\label{StructF0}
f_{k,l}^{(\infty)} \;=\;  (1-(-)^{k+l}q^{-2L})\,
\frac {(q+q^{-1})^2 \,
       \sin{\frac{\pi k}{L}} \sin{\frac{\pi l}{L}}
       \sin{\frac{\pi (k+l)}{2 L}} \sin{\frac{\pi (k-l)}{2 L}} }
      {(q^2+q^{-2}-2 \cos{\frac{\pi (k+l)}{L}}) \,
       (q^2+q^{-2}-2 \cos{\frac{\pi (k-l)}{L}}) }\,.
\end{equation}
For $q \rightarrow 1$(symmetric diffusion) this expression reduces to Eq.
(\ref{DerridaStructureFunction}).
The other two parts depend on the input rates $p_L, p_R$ and read
\begin{eqnarray}
\label{StructFL}
f^{(L)}_{k,l} &=& \frac
{\sin\frac{\pi k}{L} \, \sin\frac{\pi l}{L} \,
( \cos\frac{\pi l}{L} -\cos\frac{\pi k}{L} )}
{2\,(q+q^{-1}+\inp_L-2\cos\frac{\pi k}{L})\,
    (q+q^{-1}+\inp_L-2\cos\frac{\pi l}{L})} \\[2mm]
\label{StructFR}
f^{(R)}_{k,l} &=& \frac
{(-)^{k+l+1}\,q^{-2L}\,\sin\frac{\pi k}{L} \, \sin\frac{\pi l}{L} \,
( \cos\frac{\pi l}{L} -\cos\frac{\pi k}{L} )}
{2\,(q+q^{-1}+\inp_R-2\cos\frac{\pi k}{L})\,
    (q+q^{-1}+\inp_R-2\cos\frac{\pi l}{L})}
\end{eqnarray}
The inhomogeneous solution of the difference equations
(\ref{ScaledBulk})-(\ref{ScaledEquationsEnd}) is then obtained by
inserting Eq. (\ref{ThreeParts}) into Eq. (\ref{GeneralExpression}).

For a fixed value of the lattice length $L$ the holes probability function
depends on three parameters: $\,p_L$, $\,p_R$ and $q$. By reversing the ends
of the lattice one sees that:
\be
\Omega(j\,,m\,,p_L\,,p_R\,,q)\;=\;\Omega(L-m+1\,,L-j+1\,,p_R\,,p_L\,,q^{-1})
\label{omegpar}
\ee
Due to (\ref{ThreeParts}) the holes probability function obeys the 
following rule:
\be
\Omega(j\,,m\,,p_L\,,p_R\,,q)\;=\;\Omega(j\,,m\,,p_L\,,0\,,q)\;+\;
\Omega(L-m+1\,,L-j+1\,,p_R\,,0\,,q^{-1})\;-\;\Omega(j\,,m\,,0\,,0\,,q)
\label{TwoEndsOne}
\ee
Therefore it will be sufficient to 
study systems with particle input at only one boundary.
Using (\ref{TwoEndsOne}) one can relate physical quantities referring to 
systems with particle input at both ends with the ones computed for systems
for which $p_L$ or $p_R$ is $0$. As an example, the particle concentration
at site $j$
\be
\label{DefPartConc}
\conc(j)\;=\;1\;-\;\Omega(j-1\,,j)
\ee
can be written as a sum (\ref{TwoEndsOne}):
\be
\conc(j\,,p_L\,,p_R\,,q)\;=\;\conc(j\,,p_L\,,0\,,q)\;+\;\conc(L-j+1\,,p_R
\,,0\,, q^{-1})\;-\;\conc(j\,,0\,,0\,,q)
\label{ConcTwoEndsOne}
\ee
Here $\conc(j,0,0,q)$ is the particle concentration in the stationary state 
for input rates $0$, i.e. it is the particle concentration of one random walker
occupying the whole lattice (\ref{RW}).

%
%
%----------------------------------------------------------------------
% 4) The thermodynamic limit
%----------------------------------------------------------------------
%
\subsection{The thermodynamic limit}
\label{ThermoLimit}
\noindent
The formulas derived in the last Section
are exact solutions for finite chains. 
We consider the thermodynamical limit.
In this limit the right
boundary is moved to infinity while the observer stays in a fixed distance
$j$ to the left boundary. 
We consider systems with no particle input at the right end ($p_R\,=\,0\,$).
We are left with only two parameters,
namely the input rate at the left boundary $\inp \equiv \inp_L$ and the
asymmetry parameter $q$. Carrying out the limit
$L \rightarrow \infty$ in Eqs. (\ref{ThreeParts})-(\ref{StructFR})
one is led to a simple integral representation of the one-hole
probabilities $\Omega(j,m)$. To this end it is convenient to
introduce the quantities $\disp_z$ 
\begin{equation}
\label{TheMuFormula}
\disp_z \;=\; \frac12(q+q^{-1}-iz)-
\sqrt{\frac14(q+q^{-1}-iz)^2-1}
\end{equation}
and its inverse
\begin{equation}
\label{TheMuFormulaInv}
\disp_z^{-1} \;=\; \frac12(q+q^{-1}-iz)+
\sqrt{\frac14(q+q^{-1}-iz)^2-1}\,.
\end{equation}
Using this notation, the one-hole 
probabilities in the thermodynamic limit are given by the elliptic integral
\begin{equation}
\label{ExactIntegralRepresentation}
\Omega(j,m) \;=\; 1-\frac{q^{j+m}}{2 \pi i} \,
\int_{-\infty}^{+\infty} dz\, \Bigl( \frac{1}{z} - \frac{z}{z^2+p^2} \Bigr)
\Bigl( \disp_z^j \disp_{-z}^{m} - \disp_z^{m} \disp_{-z}^{j} \Bigr) \,.
\end{equation}
A proof of this formula is given in Appendix A where also the
particle concentration at the first few sites for infinite input rate is
computed exactly. Let us now investigate the asymptotic behavior of
the particle concentration
\begin{equation}
\label{ExactC}
c(j) \;=\;
\frac{1}{2 \pi i\,q} \,
\int_{-\infty}^{+\infty} dz\,
\Bigl(q^2\disp_z \disp_{-z}\Bigr)^{j} \,
\Bigl( \frac{1}{z} - \frac{z}{z^2+p^2} \Bigr) \,
\Bigl( \disp_z^{-1} - \disp_{-z}^{-1} \Bigr)
\end{equation}
for large $j$. Three cases have to be considered separately:
\\[5mm]
\noindent
{\bf i) Symmetric case ($q=1$):} \\[1mm] \noindent
%============================================
For $q=1$ the logarithm of the expression $\disp_z \disp_{-z}$ in
Eq. (\ref{ExactC}) can be expanded to first order in $z$ by
\begin{equation}
\log(\disp_z \disp_{-z}) \;=\; -\sqrt{2|z|}+O(|z|^{3/2})
\end{equation}
Rewriting the integral (\ref{ExactC}) by
\begin{eqnarray}
\conc(j)  &=& \frac{1}{2\pi i} \, \int_{-\infty}^{+\infty} dz\, 
\exp\Bigl(-\sqrt{2|z|}\,\, j \Bigr) \, r(j,z) \\
r(j,z) &=& \exp(\sqrt{2|z|} \,\, j)\,
(\disp_z \disp_{-z})^{j}
\Bigl( \frac{1}{z} - \frac{z}{z^2+p^2} \Bigr)\,
\Bigl( \disp_z^{-1} - \disp_{-z}^{-1} \Bigr)
\end{eqnarray}
and expanding $r(j,z)$ in $z$ the integral can be solved
order by order. We obtain the series
\begin{equation}
\label{OneOverXDecay}
\conc(j) \;=\; \frac{2}{\pi j}+\frac{1}{\pi j^2} -\frac{1}{2 \pi j^4} 
+\Biggl(\frac{3}{8 \pi} - \frac{12}{\pi \inp^2}\Biggr)\frac 1 {j^5} + O(j^{-6})
\end{equation}
This proves that in the fermionic case  the first three terms in the large $x$
expansion are independent of the input rate $\inp$.
\\[5mm]
\noindent
{\bf ii) Bias to the right ($q>1$):} \\[1mm] \noindent 
%============================================
In this case we find that the expression $\log(q^2\disp_z \disp_{-z})$ 
in Eq. (\ref{ExactC}) can be expanded in first order by:
\begin{equation}
\log(q^2\disp_z \disp_{-z}) \;=\; 
-q^2 \frac{q^2+1}{(q^2-1)^{3}} \,z^2 -
O(z^4) \,.
\end{equation}
Rewriting the integral (\ref{ExactC}) by
\begin{eqnarray}
\conc(j)  &=& \frac{1}{2 \pi i\, q} \, \int_{-\infty}^{+\infty} dz\, 
\exp\Bigl(-q^2 \frac{q^2+1}{(q^2-1)^{3}} \, j \, z^2 \Bigr) \, s(j,z) \\
s(j,z) &=& \exp\Bigl(q^2 \frac{q^2+1}{(q^2-1)^{3}} \,j \, z^2 \Bigr)
\Bigl( \frac{1}{z} - \frac{z}{z^2+p^2} \Bigr)
(q^2\disp_z \disp_{-z})^{j} \Bigl( \disp_z^{-1} - \disp_{-z}^{-1} \Bigr)
\end{eqnarray}
and expanding $s(j,z)$ in $z$ the integral can be solved again
order by order. We obtain
\begin{equation}
\label{AsymmetricProfile}
\conc(j) \;=\;  \sqrt{\frac{q^2-1}{(q^2+1)\pi j}}
\,\Biggl[ 1 + \Biggl(\frac{3q^4+20q^2-1}{8(q^4-1)} -
\frac{(q^2-1)^3}{2q^2(q^2+1)\inp^2 }\Biggr)\frac 1 j \Biggr] + O(j^{-5/2})
\end{equation}
We notice that, as opposed to the symmetric case, only the leading term is 
independent of the input rate.

\noindent
{\bf iii) Bias to the left $q<1$:} \\[1mm] \noindent
If the particles hop preferentially to the left, they accumulate at
the left boundary and thus we expect an exponential decay of the
concentration profile. In fact, as can be seen from Eq. (\ref{ExactC}),
the expression $q^{2j-1} \conc(j)$ is invariant under the replacement
$q \rightarrow q^{-1}$. This means that for a bias directed towards the left
boundary the concentration profile decays like $q^{4j}j^{-1/2}$:
\begin{equation}
\conc(j) \;=\;  q^{4j} \left(
\sqrt{\frac{1-q^2}{(q^2+1)\pi j}} + O(j^{-3/2}) \right)
\end{equation}

We would like to remark that the series presented in this Section
are {\it asymptotic series} 
since they are derived from an elliptic integral.
%
%
%
%
%----------------------------------------------------------------------
% The continuum limit
%----------------------------------------------------------------------
%
\subsection{The continuum limit}
\label{ContinuumSection}
An alternative way to describe the physics of
%and more intuitive 
the coagulation model with an external input source is to consider
the continuum limit of the one-hole equations.
This can be done
by taking the lattice spacing $\spacing \rightarrow 0$ \
while keeping the two quantities
\be
\label{defcv}
\pac  = \frac{2\,(q-q^{-1})} {\spacing\,(q+q^{-1})}\,
\hspace{10mm} \mbox{and} \hspace{10mm}
\hat{\inp} = \frac {2 \inp_L} {\spacing^2(q+q^{-1})}
\ee
constant. We then replace the empty-hole probabilities
$\Omega(x,y)$ by their continuous counterparts:
\be
\Omega^{c}(x,y) \;=\;
\Omega(\frac{j}{\spacing},\frac{m}{\spacing})
%\Omega(\frac{x}{\spacing},\frac{y}{\spacing})\,.
\ee
It is useful to rescale the hole density function taking ${\hat{\Omega}}(x,y) =
\Omega^{c}(x,y)e^{-\frac{\pac }{2}(x+y)}\;$ which verifies the equation (see
Eq.(\ref{ScaledBulk}))
\ba
\label{ContinuousBulk}
\Bigl( \Delta - \frac{\pac ^2}{2} \Bigr) \, \hat{\Omega}(x,y) &=& 0
\hspace{15mm} (L>y>x>0)
\ea
By solving the continuous counterparts of Equations 
(\ref{LeftBoundaryCondition})-(\ref{FullEquationsEnd}) 
we determined the value 
of the holes density function on
the boundaries. The solutions are:
\begin{itemize}
\item Along the left boundary ($x=0\;,\;0\,\leq\,y\,\leq\,L$):
\begin{eqnarray}
\label{CL}
\hat{\Omega}(0,y) &=& \left\{
\begin{array}{cc}
	\frac{\sinh{\Bigl((L-y)\;\sqrt{\frac{\pac ^{2}}{4}+\hat{p}}}
	 \Bigr)} {\sinh{\Bigl(L\;\sqrt{\frac{\pac ^{2}}{4}+
	\hat{p}}}\Bigr)} 
 &\mbox{if \ } \hat{p}\;\neq \infty\\
0&\mbox{if \ } \hat{p}\; =   \infty 
\end{array} \right. \\ \nonumber
\end{eqnarray}

\item Along the upper boundary ($y=L\;,\;0\,\leq\,x\,\leq\,L$):
\begin{eqnarray}
\label{CR}
\hat{\Omega}(x,L) &=& \left\{
\begin{array}{cc}
	\frac{e^{-\pac  L}\sinh{\Bigl(\frac{\pac }{2}x}\Bigr)}
	{\Bigl(\sinh{\frac{\pac }{2}L}\Bigr)} 
 &\mbox{if \ } \pac \; \neq 0 \\
x/L  & \mbox{if \ } \pac \; = 0 
\end{array} \right.
\end{eqnarray}
\item On the diagonal ($0\,\leq\,x\,=\,y,\leq\,L\,$) the normalisation 
condition is:
\begin{equation}
\label{CD}
\hat{\Omega}(x,x) \;=\; e^{-\pac  x}\,.  
%\hspace{15mm} (L \geq x \geq 0)
\end{equation}
\end{itemize}
Equation (\ref{ContinuousBulk}) together with the boundary conditions
(\ref{CL})-(\ref{CD}) define a Dirichlet problem for the function
$\hat{\Omega}(x,y)\,$. The formal solution is
\be
\label{ContourIntegral}
\hat{\Omega}(x,y)  \;=\; \oint_C ds\, \hat{\Omega}(x',y')
\frac{\partial}{\partial n} \, {\cal G}(x,y,x',y')
\ee
where $C$ is the contour along the boundaries, $\frac{\partial}{\partial n}$
the normal derivative and ${\cal G}(x,y,x',y')$ the Green function defined by
\begin{eqnarray}
\label{defgr}
&&\Bigl( \Delta - \frac{\pac ^2}{2} \Bigr){\cal G}(x,y,x',y') = 
\delta(x-x')\delta(y-y')\\
&&{\cal G}(0,y,x',y')= {\cal G}(x,y,0,y')=
  {\cal G}(x,L,x',y')= {\cal G}(x,y,x',L)=0\\
&&{\cal G}(x,x,x',y')= {\cal G}(x,y,x',x')= 0
\end{eqnarray}
The computation of the density function of the hole probabilities requires the 
computation of the Green function. This was done in two steps. First
notice that:
\be
\nonumber
{\cal G}(x,y,x',y')={\cal G}_{\Box}(x,y,x',y')-{\cal G}_{\Box}(y,x,x',y')
\ee
where ${\cal G}_{\Box}(x,y,x',y')$ is the Green function of the Dirichlet 
problem defined in the interior of the square $\,0 \leq x \leq L\,$,
$\,0 \leq y \leq L$. $G_{\Box}$ can be constructed
easily by using reflection techniques.  All what one needs to know is
the Green function of the Dirichlet problem defined on the entire plane
(with boundaries at infinity). We denote the last mentioned function with 
$g(x,y,x',y')$. Summing up, we get:
\begin{eqnarray}
\label{Green}
{\cal G}(x,y,x',y') &=& \sum_{\alpha,\beta=\,\pm 1} \alpha\beta
\sum_{i,j=-\infty}^{+\infty} \Biggl[ g\,\Bigl(\sqrt{(x-2iL-\alpha x')^2\,+\,
(y-2jL-\beta y')^2}\Bigr)\;-\; \\ \nonumber && \hspace{30mm}
g\,\Bigl( \sqrt{(y-2iL-\alpha x')^2\,+\,
(x-2jL-\beta y')^2}\Bigr) \Biggr]
\end{eqnarray}
Once the density function $\hat{\Omega}(x,y)$ is known one can compute 
expectation values of observables in the steady state.
The local particle density is given by
\be
\label{LocalDensity}
\density(x) \;=\; \lim_{y \rightarrow x}
\frac{1-e^{\frac{\pac }{2}(x+y)} \, \hat{\Omega}(x,y)} {y-x}  \;=\;
- \left. \frac{\partial}{\partial y}\,
\Omega^{c}(x,y) \right|_{y=x}\,.
\ee
We also give a closed formula for the computation of the connected 
two-point function:
\be
G^{c}(x,y) = \langle \dens_x \dens_y \rangle -  \langle \dens_x \rangle
\langle \dens_y \rangle
\label{defc2p}
\ee
Here $\dens_x$ denotes the particle number operator at site $x$.
Using the factorization properties of the two-holes probability function
mentioned in \cite{Peschel, PointlikeSourceExact} 
it is easy to see that in the continuum limit we have:
\be
G^{c}(x,y) = \frac{\partial}{\partial x} \Omega^{c} (x,y) \frac{\partial}
{\partial y} \Omega^{c} (x,y) - \Omega^{c} (x,y) \frac{\partial^2 \Omega^{c}
(x,y)} {\partial x \partial y}
\label{TPF}
\ee

\subsubsection{The scaling limit in the symmetric case ($q=1$)}
\label{scalinglimit}
%
%
%==================================================
In the case of symmetric diffusion the differential equation 
(\ref{ContinuousBulk}) reduces to a Laplace equation. The Green function
can be obtained from (\ref{Green}) by replacing $g(u)$ with $\frac{1}{2 \pi} 
\ln u$.

Let us consider for simplicity first the case of an {\em infinite particle 
input} rate ($\hat{p}= \infty$).  In this case the hole density function is 
zero for $x=0\;$ (see Eq. (\ref{CL})). We are left with:
\be
\Omega^{c}_{\infty}(x,y) \;=\; \int_0^L du\, \left.
\left( \frac{\partial}{\partial x'} - \frac{\partial}{\partial y'} \right)
\, {\cal G}(x,y,x',y') \, \right|_{x'=y'=u}
\;-\; \int_0^L dx'\, \left.  \frac{x'}{L} \frac{\partial}{\partial x'}
{\cal G}(x,y,x',y') \, \right|_{y'=L}
\ee
Inserting Eq. (\ref{Green}) one is led to
\begin{eqnarray}
\label{Oq1pi}
&& \Omega^{c}_{\infty}(x,y) \;= \\ \nonumber
&& \sum_{\alpha,\beta=\,\pm 1} \alpha\beta
\sum_{i,j=-\infty}^{+\infty} \Bigl[ \arctan \Bigl( \frac
{\alpha (2i\,-\,x/L) - \beta (2j\,-\,y/L)}{(2i\,-\,x/L)^2+(2j\,-\,y/L)^2
+\alpha (2i\,-\,x/L) + \beta (2j\,-\,y/L)}\Bigr) \nonumber \\
&& \nonumber \hspace{30mm}
-\Bigl(\beta (i\,-\,\frac{x}{2L}) + \alpha (j\,-\,\frac{y}{2L})
\Bigr) \arctan \frac {\alpha\,+\,2i\,-\,x/L}{\beta\,+\,2j\,-\,y/L} \Bigr]\,.
\end{eqnarray}
We see that the hole density function for $\hat{p}=\infty\;$ depends only
on $\;x/L\;$ and $\;y/L\;$. From 
Eq. (\ref{LocalDensity}) we obtain the local particle density in
the stationary state for {\it infinite} particle input rate:
\be
\density_{\infty}(x) \;=\; \frac{2}{\pi L}\,\sum_{i,j=-\infty}^{+\infty}
\frac{(x/L-2i)+(x/L-2j)}{(x/L-2i)^2+(x/L-2j)^2}
\label{Cq1pi}
\ee
Defining $z=\frac x L$, Eq.(\ref{Cq1pi}) can be rewritten as
%An important result is that the one-point-function scales like
%
%
\be
L\density_{\infty}(x) \;=\;\Phi(z)
\label{dsl}
\ee
where
\ba
\label{SFQ1}
\Phi(z)&=&\frac{1}{\pi}\,
\sum_{i,j=-\infty}^{+\infty}\frac{(z/2-i)+(z/2-j)}{(z/2-i)^2+(z/2-j)^2}\;=
\\ \nonumber
 &=&\; \frac{\sinh(\pi z)\,+\,
\sin(\pi z)}{\cosh(\pi z)\,-\,\cos(\pi z)}\;+\;
\sum_{k=1}^{+\infty}\Biggl\{\frac{\sinh(2\pi(z/2-k))\,+\,
\sin(\pi z)}{\cosh(2\pi(z/2-k))\,-\,\cos(\pi z)}\;\;+\;(\;k\;
\longleftrightarrow \;-k\;)\Biggr\}\,.
\ea
The function $\,\Phi(z),\,$ called scaling function, is odd and periodic
with period 2.
In the limit $\;z\rightarrow 0\;$ the function diverges like $\;\frac
{2}{\pi z}\;$ (the dominant contribution is given by the first term
in the second line of (\ref{SFQ1})). For 
$\;z\rightarrow 1,\; \Phi(z)$ approaches the value 1, 
%the scaling function 
but  the value itself in the point $\,z\,=\,1\,$ is $\;\Phi(1)\,=\,0\;$. 
So the function is discontinuous for all integer arguments.

We consider now the case of an arbitrary input rate $\hat{p}$ and look at 
the scaling regime ($z$ fixed,$\;L\rightarrow\,\infty$).
If $\hat{p}\;$ is finite one picks up another contribution to the holes density
function coming from the integration along the boundary segment ($\,x\,=0 
\; , \;\;0\,\leq y\leq L$) in (\ref{ContourIntegral}). The difference between 
the values of the holes density function corresponding to infinite and finite 
input rates is:
\be
\label{CorrectionIntegral}
\Omega_\infty^c (x,y) - \Omega_{\hat{p}}^c(x,y) \;=\;
\int_0^L dy'\,\left.\Omega_{\hat{p}}^c(0,y') \,
\frac{\partial}{\partial x'} {\cal G}(x,y,x',y')
\right|_{x'=0}
\ee
We are here interested in the scaling and thermodynamical limit.
For large values of the lattice length $L\;$, $\Omega_{\hat{p}}^c(0,y')$
behaves like $\;\exp(-y' \sqrt{\hat{p}})\;$. We expand the
derivative of the Green function in Eq. (\ref{CorrectionIntegral})
near $\;y'\,=\,0\;$ and get:
\ba
\Omega_\infty^c (x,y) - \Omega_{\hat{p}}^c (x,y) &=&
\frac{16}{\pi}\; \sum_{i,j=-\infty}^{+\infty}\frac{(x/L-2i)(y/L-2j)
\Bigl[(y/L-2j)^2\,-\,(x/L-2i)^2\Bigr]}
{\Bigl[(y/L-2j)^2\,+\,(x/L-2i)^2\Bigr]^4}
 \times \nonumber \\ 
 & &  \int_0^1  \frac{\sinh{(\sqrt{\hat{p}}L(1-u))}}
{\sinh{(\sqrt{\hat{p}}\;L})}\Bigl(\,u^3\, +\;O(u^7)\,\Bigr)  du
\nonumber
\ea
It is easy to see that in the finite-size scaling limit one gets:
\be
\Phi_{\hat{p}}(z)\;=\;\Phi(z)\,+\,\frac{1}{L^4}\Phi^{\mbox{\small corr}}(z) 
+O\Bigl(\frac{1}{L^8}\Bigr)\,.
\label{scalcomp}
\ee
The first finite-size scaling correction function is
\be
\Phi^{\mbox{\small corr}}(z)\;=\;\frac{3}{\pi\hat{p}^2}
\sum_{i,j=-\infty}^{+\infty}(z/2-i)\frac{(z/2-i)^4-10(z/2-i)^2(z/2-j)^2
+5(z/2-j)^4}{\Bigl[(z/2-i)^2+(z/2-j)^2\Bigr]^5}
\label{corrfunc}
\ee
So we proved that in the scaling limit the particle density
function scales. The scaling function $\;\Phi(z)\;$ given by equation 
(\ref{SFQ1}) is independent on the input rate. We get finite-size corrections
of order $\;L^{-4}\;$ for finite input rates. 

In Figure \ref{f1} we show the function (solid curve) 
\be
F(z)\,=\,\frac{\pi z}{2}\cdot \Phi(z)
\label{rsf}
\ee
(with this definition $F(0)=1$) together with finite-lattice calculations
(see Eqs. (\ref{DefPartConc}) and (\ref{GeneralExpression})) obtained for
$L=2000$ ($p=1$) and for $L=800$ ($p=\infty$). We have given this figure
for two reasons. First we observe that the scaling function has a nontrivial
behavior at the opened end of the system ($z\propto 1$).  Next, although
the scaling function was computed in the continuum limit, it applies to the
lattice too.

\vspace{3mm}
Let us consider the {\em thermodynamical limit} $ L\rightarrow \infty,\;\;x 
\;\; \mbox{fixed}$. Using equations (\ref{LocalDensity}) and (\ref{Green}) 
(from the multiple sum of the latter one we take only the terms corresponding 
to $i\,=\,j\,=\,0$) one can show that
\be
\label{Correction}
\density(x) \;=\;
\frac 2  {\pi x}\,-\,\frac{12}{\pi \hat{p}^2 x^5} +
\frac{5040}{\pi \hat{p}^4 x^9} + O(p^{-6}x^{-13})\,.
\ee
The asymptotic behavior (for large values of $x$) of the one-point function
in the thermodynamical limit (\ref{Correction}) can be also obtained from
the scaling behavior (\ref{scalcomp}), in the limit $\;z\rightarrow 0\;$. 
The result (\ref{Correction}) is consistent
with the expansion for the stationary concentration profile on
a discrete lattice (\ref{OneOverXDecay}). The only difference is that
in the continuum limit all contributions $\frac{1}{x^ip^j}$ with
$i>2j+1$ scale like $\lambda^{i-2j-1}$ and therefore vanish in the limit
of vanishing lattice spacing $\lambda \rightarrow 0$.

We mention one last result concerning the connected two-point function.
In the thermodynamical limit one can see 
(c.f. Ref. \cite{PointlikeSourceExact})  that 
the hole probability density for infinite input rate is:
\be
\Omega^c_{\infty}(x,y)\;=\;\frac{4}{\pi} \arctan \Bigl(\frac{x}{y}\Bigr)
\label{4op}  
\ee
%
%
%This is given by the term ($\,i\,=\,j\,=\,0\,$) in the sum (\ref{Oq1pi}).  
Using (\ref{TPF}) one gets the following expression for the connected 
two-point-function
\be
G^c (x,x+d)\;=\;-\frac{16}{\pi^2}\frac 1 {x^2} 
\frac{\Bigl[1+v-v(2+v)\,\arctan\Bigl(\frac{1}{1+v}\Bigr)\Bigr]}
{(2+2v+v^2)^2}
\label{CTPF}
\ee
in the infinite input rate case. On the right hand side of (\ref{CTPF})
we denoted $\frac d x$ with $v$. Notice again the algebraic fall-off.
\noindent
%\\[5mm]
%
%

\subsubsection{Bias to the right ($\pac >0$)}
\label{biasdiffcon}

Writing the Fourier transform of (\ref{defgr}) one gets
\be
g(u)\;=\;
-\frac{1}{(2 \pi)^2} \int_0^\infty \frac{ k \cdot dk}{ k^2 + \pac ^2/2}
\,\int_0^{2 \pi} e^{-ik u \cos{\theta}} d\theta\,.
\ee
Thus the Green function of the problem defined on the entire plane is
(see formulae (9.6.16) in \cite{Abr} and (6.532.4) in \cite{Gra})
\be
\label{K0}
g(u)\;=\;
-\frac{1}{2 \pi}\,K_0\Bigl( \sqrt{\frac{\pac ^2}{2}}u\Bigr)
\ee
We use the standard notation $K_i\,,\,\,i=0,1,2 ...$ for the modified Bessel 
functions.

In Appendix B we will prove that in the thermodynamical limit the 
asymptotic behavior of the particle density is given by
\be
\label{qsf1}
\density (x)\;=\;\sqrt{\frac \pac  {2\pi x}} \,+\,\frac 1 {4\sqrt{2\pi \pac }}
\Biggl[ \frac {11} 2 \,-\,\frac{\pac ^4}{\hat{p}^2} \Biggr]
\frac 1 {x^{3/2}}\;+\;O(x^{-5/2})
\ee
The leading term is the continuum correspondent of the one appearing in
Eq. (\ref{AsymmetricProfile}) and is independent of the input rate.

\vspace{0.5cm}

We conclude this Section with some remarks concerning the influence of the
right boundary on the particle density.

As opposed to the symmetric case, one can check that for biased diffusion to 
the right the behavior of the one point function in the thermodynamical limit 
is identical with the one in the scaling limit. 
One can give a qualitative explanation. In the stationary state, near the
right boundary, the density function can be approximated with the one given by 
a single particle occupying the whole lattice (a random walker)
\begin{eqnarray}
\label{RW}
\conc(y) &=& \left\{
\begin{array}{cc}
		q^{-2y}\frac{1-q^{-2}}{1-q^{-2L}}
 &\mbox{for } q\; \neq \; 1\\
 & \\
\frac 1 L &\mbox{for } q\; = \;1
\end{array} \right.
\end{eqnarray}
where $y=L-x$.

For biased diffusion to the right the density decays exponentially in $y$. 
So the influence of the right boundary is of short range and is not seen in the
scaling limit.

For $q=1$ (or $\pac =0$) the influence of the right boundary is much stronger.
The density determined by a random walker near the right end of the lattice 
is constant ($\frac 1 L$) and greater than the one obtained through
the extrapolation of Eq. 
(\ref{OneOverXDecay}) ($\frac 2 {\pi L}$). This explains
the linear behavior of the reduced scaling function for $z\,\rightarrow \,1$ 
(see Figure \ref{f1}).

One can also notice from the $y$ and $L$ dependence of $c(y)$ that one has a
characteristic length scale $\Lambda\,=\,(2\ln q )^{-1}$ which is related
to the inverse mass seen in the spectrum of the Hamiltonian (see Eq.
(\ref{tferm})). This observation is very interesting since it clarifies a 
puzzle which 
goes through this paper: how can a system with massive excitations in the time
direction show an algebraic and, as we shall see in the next Section, universal
behavior? The answer is that one looks at the concentration in the "wrong"
way following the $x$ dependence (away from the source) and not the $y$ 
dependence (away from the open end). The discovery that this "wrong" way
exists is probably the main achievement of this paper.
%
%----------------------------------------------------------------------
%  Results from Monte Carlo data
%----------------------------------------------------------------------
%
\section{Numerical verification of the universality hypothesis}
\setcounter{equation}{0}
\label{Numerical results}

In this Section we present the results of Monte Carlo simulations.
The details are given in Appendix~C.
We restrict the study to cases for which $p_R=0$ and $\pa\,\geq\,0$.
We start with the case of symmetric diffusion ($q=1$ or $\pa\,=\,0$).
First we look at the density profile ($1\,\ll\,x\,\ll\,L$). As suggested
by Eq.(\ref{OneOverXDecay}), we fit the data by the function
\be
\label{fitsq1}
c(x)\,=\,\frac{K_1}{x} \,+\,\frac{K_2}{x^2}\,+\,\frac{K_3}{x^3}\,.
\ee
%
%
%Using the $\chi^2$ method we estimated the coefficients $K_1,\ldots,K_3$.
%In order to avoid finite-size effects, only points in the interval
%$[\,L/10\;,\;L/2\,]$ are taken into account.
%The estimates for $K_1$ and $K_2$ for various input and bulk rates are 
%given in Table \ref{tabq1}. The data presented here is obtained for the
%lattice size $L=1000$.
When making the fits (by using the $\chi^2$ method) we took points
$\,x\,\in\,[\,L/10\;,\;L/2\,]$ in order to avoid finite-size effects.
The estimates for $K_1$ and $K_2$ for various input and bulk rates are
given in Table \ref{tabq1}. The data presented here were obtained taking
lattices of size $L=1000$.

\begin{table}[htb]
\centering
\begin{tabular}{|c|c|c||c|c|} \hline
\multicolumn{1}{|c|}{$\pb$} &
\multicolumn{1}{|c|}{$\pc$} &
\multicolumn{1}{|c|}{$p$} &
\multicolumn{1}{c|}{$K_1$}&
\multicolumn{1}{c|}{$K_2$} \\ \hline \hline
$ 0.50 $&$ 0  $&$  1      $&$ 0.635 \pm 0.001$&$ 10.2 \pm 0.2$\\
$ 0.50 $&$ 0  $&$  \infty $&$ 0.639 \pm 0.004$&$  9.8 \pm 0.8$\\
$ 0.50 $&$-0.2$&$  1      $&$ 0.640 \pm 0.010$&$ 16   \pm 5  $\\
$ 0.50 $&$-0.2$&$  \infty $&$ 0.640 \pm 0.010$&$ 13   \pm 3  $\\
$ 0.25 $&$ 0  $&$  1      $&$ 0.637 \pm 0.002$&$ 22   \pm 4  $\\
$ 0.25 $&$ 0  $&$  \infty $&$ 0.632 \pm 0.002$&$ 24.4 \pm 0.4$\\
$ 0.40 $&$-0.2$&$  1      $&$ 0.638 \pm 0.002$&$ 11   \pm 2  $\\
$ 0.40 $&$-0.2$&$  \infty $&$ 0.642 \pm 0.004$&$ 10.4 \pm 0.6$\\
\hline
\end{tabular}
\caption
[Estimates of the coefficients $K_1$ and $K_2$ of the expression
(\protect\ref{fitsq1}) for various input and bulk rates]
{Estimates of the coefficients $K_1$ and $K_2$ of the expression
(\protect\ref{fitsq1}) for various input and bulk rates}
\label{tabq1}
\end{table}

We notice that $K_1$ is everywhere close to the value obtained in the
fermionic case ($\pb = 2,\;\pc=0$) namely
$\,K_1\,=\,2/\pi\,\simeq \, 0.637$. The values of $K_2$ are different if the 
bulk rates are different but as in the fermionic case they do not depend
on the input rate.

Since the leading term of the density profile is compatible with 
universality, one can go one step further and check if the scaling function
$\Phi(z)$ given by Eq. (\ref{SFQ1}) in the fermionic case, is also universal.
This function was obtained taking $x/L\,=\,z\,$ fixed ($x$ and $L$ large):
\be
\lim_{L\rightarrow \infty}L c(z,L) \;=\;\Phi(z)
\ee
We define the function
\be
\label{kzl}
K(z,L)\;=\;\frac{L\;c(z,L)}{\Phi(z)} -1
\ee
which measures the deviation from universality and the finite-size effects.
In Fig. \ref{f2} we give the data in the case $p=1\,,\pb=0.5\,,\pc=0\;$ for 
three lattice sizes. One notices that with increasing lattice size $K(z,L)$ 
decreases, as it should. One should mention that for $z=0$ one expects
$K(z,L)$ to go to zero in the limit $L \rightarrow \infty$ because of the 
universality of $K_1$ in Eq.(\ref{fitsq1}) and the uniform convergence
of $\Phi(z)$ in the fermionic case. Thus the relative large values
of $K(z,L)$ observable in Fig.~\ref{f2} for small values of $z$ should not be a 
subject of concern.
We have also done other simulations (not shown in Figure \ref{f2}) for 
other input rates which show the same properties.

In Figure \ref{f3}, $K(z,L)$ is shown for various input and coagulation rates
for a lattice of length $L=1000$. As one can see $K(z,L)$ is small everywhere
(for small values of $z$ the convergence is slow but as mentioned above,
for $z=0$ universality was checked already).

We now consider the asymmetric diffusion case ($q>1$). As suggested by Eq.
(\ref{AsymmetricProfile}) we fit the Monte Carlo data by the function
\be
\label{fitsq5}
c(x)\,=\,K'_{1/2}x^{-1/2} \,+\,K'_1 x^{-1}\,+\,K'_{3/2}x^{-3/2}\,.
\ee

We choose $\,q\,=\,\sqrt{2.5} \;$ (which corresponds to $\pa=0.857$),
in which case we get from Eq. (\ref{AsymmetricProfile}) $K'_{1/2}\,=\,
\sqrt{\frac 3 {7 \pi}}\,\simeq\,0.369$. Notice that we have allowed a 
term $\sim x^{-1}\,$ not present in Eq. (\ref{AsymmetricProfile}).
The data were collected for $L=1000$ and the results are shown in Table 
\ref{Tabq5}.

\begin{table}[htb]
\centering
\begin{tabular}{|c|c|c||c|c|} \hline
\multicolumn{1}{|c|}{$\pb$} &
\multicolumn{1}{|c|}{$\pc$} &
\multicolumn{1}{|c|}{$p$} &
\multicolumn{1}{c|}{$K'_{1/2}$} &
\multicolumn{1}{c|}{$K'_1$} \\ \hline \hline
$1.000$&$ 0.429$&$  1     $&$ 0.370 \pm 0.003$&$ 0.10 \pm 0.01$\\
$1.000$&$ 0.429$&$ \infty $&$ 0.368 \pm 0.002$&$ 0.11 \pm 0.01$\\
$0.500$&$ 0.214$&$  1     $&$ 0.368 \pm 0.003$&$ 0.90 \pm 0.04$\\
$0.500$&$ 0.214$&$ \infty $&$ 0.369 \pm 0.004$&$ 0.82 \pm 0.07$\\
$0.250$&$ 0.107$&$  1     $&$ 0.369 \pm 0.002$&$ 2.12 \pm 0.02$\\
$0.250$&$ 0.107$&$  \infty$&$ 0.369 \pm 0.002$&$ 2.09 \pm 0.02$\\
$0.361$&$-0.181$&$   1    $&$ 0.368 \pm 0.001$&$ 1.27 \pm 0.02$\\
$0.361$&$-0.181$&$ \infty $&$ 0.369 \pm 0.003$&$ 1.25 \pm 0.04$\\
\hline
\end{tabular}
\caption
[The coefficient $K'_{1/2}$ and $K'_1$ of the expansion  
(\protect\ref{fitsq5}) for various input and coagulation rates. All data 
are for $q= \protect\sqrt{2.5}\; (\protect\pa = 0.857)$.]
{The coefficient $K'_{1/2}$ and $K'_1$ of the expansion  
(\protect\ref{fitsq5}) for various input and coagulation rates. All data 
are for $q= \protect\sqrt{2.5}\; (\protect\pa = 0.857)$.}
\label{Tabq5}
\end{table}

As can be seen from this table the coefficient $K'_{1/2}$ is unchanged
(universal). The $K'_1$ is independent on the input, but depends on the
bulk rates (similar to the $K_2$ coefficient in~Eq. (\ref{fitsq1})).
Finally we notice that the values of $K'_1$ get smaller if we approach the
fermionic case ($\pa\,=\,\pc\,=\,0.857,\;\pb\,=\,2\;$).

To sum up, in the symmetric diffusion case the Monte Carlo data suggest
that the large $x$  behavior and the scaling function are universal: 
they are independent of the $c_L,\,c_R\,$ and $p$ rates. 
In the asymmetric diffusion case, the large $x$ behavior is also
universal.
%
%
%----------------------------------------------------------------------
% The annihilation model with external particle input and output
%----------------------------------------------------------------------
%
%
\section{Connection with other models}
\label{AnnModels}
\setcounter{equation}{0}
It is a well-known fact that the coagulation model $A+A\rightarrow A$
and the annihilation model $A+A \rightarrow \0$ belong to the same
universality class. This equivalence is due to the existence of a local
similarity transformation between their time evolution operators 
\cite{Horatiu}.

We now use this transformation in order to
apply the results of the preceding Sections to a coagulation-annihilation
model (called CA) with boundary effects which is defined by the following 
processes and rates:
$$
\begin{array}{ll}
A\0 \rightarrow \0 A &
\mbox{diffusion to the right at rate $\tilde{a}_R$} \\
\0 A \rightarrow A\0 &
\mbox{diffusion to the left at rate $\tilde{a}_L$} \\
AA \rightarrow \0 A &
\mbox{coagulation to the right at rate $\tilde{c}_R$} \\
AA \rightarrow A\0 &
\mbox{coagulation to the left at rate $\tilde{c}_L$} \\
AA \rightarrow \0\0 &
\mbox{pair annihilation at rate $\tilde{\kappa}$}
\end{array}
$$
In addition particles are absorbed (desorbed) at rate $\tilde{\gamma}$
($\tilde{\delta}$) at the left boundary. In the configuration basis
the time evolution operator
$H^{CA} \;=\; I^{CA}_1 + \sum_{n=1}^{L-1}  H^{CA}_{n,n+1}$
is given by
\begin{equation}
\label{AnnHamiltonian}
H^{CA}_{n,n+1} \;=\; \left(\begin{array}{cccc}
		   \,\,\,0\,\, & 0    & 0    & -\tilde{\kappa}    \\
		   0 & \tilde{a}_L  & -\tilde{a}_R & -\tilde{c}_R \\
		   0 & -\tilde{a}_L & \tilde{a}_R  & -\tilde{a}_L \\
		   0 & 0    & 0    & \tilde{\kappa}+\tilde{c}_R+
		\tilde{c}_L
		  \end{array} \right)\, ,
\hspace{15mm}
I^{CA}_1  \;=\; \left(\begin{array}{cc} \tilde{\gamma} & -\tilde{\delta} \\ 
	-\tilde{\gamma} & \tilde{\delta}
		  \end{array} \right)
\end{equation}
\noindent
As shown in Ref. \cite{Horatiu}, the coagulation model (\ref{Hamiltonian})
and the generalized annihilation model (\ref{AnnHamiltonian}) are related
by a local similarity transformation $H^{CA}= U H^{coag} U^{-1}$ 
\begin{equation}
\label{SimilarityTransformation}
U \;=\; u \otimes u \otimes \ldots \otimes u \;=\; u^{\otimes L} \,,
\hspace{15mm}
u \;=\; \left( \begin{array}{cc} 1&1-a\\0&a \end{array} \right)
\end{equation}
where $a$ is some parameter.
The rates of the coagulation-annihilation model are
related to those of the original coagulation model by
\begin{eqnarray}
\tilde{a}_{L,R} &=& a_{L,R} \nonumber \\
\tilde{c}_{L,R} &=& c_{L,R} + \frac{1-a}{a} (a_{R,L}-a_{L,R}-c_{R,L}) 
\nonumber \\
\tilde{\kappa} &=& \frac{1-a}{a} (c_L+c_R) \\
\tilde{\gamma} &=& a p \nonumber \\
\tilde{\delta} &=& (1-a) p \nonumber
\end{eqnarray}
Notice that if the original model had only input of particles the equivalent 
coagulation-annihilation model has both input and output of particles.

\noindent
Because of the simplicity of the transformation the $n$-point 
density-density correlation functions in 
the coagulation and coagulation-annihilation model are related by
\begin{equation}
\langle \tau_{j_1} \tau_{j_2} \ldots \tau_{j_n} \rangle^{CA} 
\;=\;a^{n} \, 
\langle \tau_{j_1} \tau_{j_2} \ldots \tau_{j_n} \rangle^{coag}
\end{equation}
%
%
%----------------------------------------------------------------------
% Conclusions
%----------------------------------------------------------------------
%
\section{Conclusions}
\label{ConclusionSection}
\setcounter{equation}{0}

\indent
In the present paper we investigated the coagulation-diffusion model with
particle input at one boundary using both analytical and numerical methods. The
results show that spatial long-range correlations play an essential role
and that some physical properties are universal with respect to the input
and the coagulation rates.

We started our analysis with a simple space-dependent mean field
approximation. It predicts algebraic behavior of the
particle density in the stationary state for both symmetric and biased
diffusion. 
However, rigorous results require an exact solution of the
problem. To this end we solved the full problem by using the IPDF
formalism. This formalism can be used only if the coagulation rates 
coincide with the diffusion rates, which corresponds to free fermions 
in the Hamiltonian language. The large $x$ behavior ($x$ is the distance 
to the source) of the particle density was computed in the thermodynamical
limit both for the lattice and the continuum version and the results are
compared. These painful calculations were done for symmetric and asymmetric
diffusion. In the case of symmetric diffusion the scaling limit ($x/L$ 
fixed, $L$ is the lattice length) was obtained.

Monte Carlo simulations show that the coefficient of the leading terms of
the asymptotic expansion of the density in the thermodynamical limit are
universal: they are independent of the input rates (this was to be expected
from mean-field) and on the
coagulation rates. The scaling function is also universal in the symmetric 
case. It is trivial in the asymmetric case (it coincides with the leading 
term of the large $x$ behavior of the density). These results were to be 
expected from common sense in the 
symmetric case but not for the asymmetric case. The reason is the following
one: the relaxation spectrum of the system is massless
in the first but massive in the second case. There exists a myth according
to which if there are lengths in the time evolution there should be lengths 
in the space correlations. A counter-example can be found however in the 
kinetic Ising model (see  Ref. \cite{Alcaraz}). In the coagulation-diffusion 
model the picture is more perverse~: if one looks at the concentration 
starting at the opened end, one finds an exponential fall-off but an
algebraic and universal behavior if we start at the source end.

The message of this paper can be extended to the problem in which we add
pair-annihilation in the bulk and an output of particles at the source.
What is still missing is a proof of universality which goes beyond numerical
checks. This can be done using field theoretical methods \`a la Cardy
\cite{CBL, CPC}.
\\[10mm]
%\noindent
{\bf Acknowledgments}\\[2mm]
We would like to thank J. Cardy, B. Derrida, K. Krebs, I. Peschel and 
G. M. Sch\"utz for helpful hints and interesting discussions and to P. A.
Pearce for reading the manuscript. H. H. would 
like to thank the Minerva foundation for financial support. H. S. would like
to express special thanks to E. Heged\"us for teaching him Green functions.

\appendix
\renewcommand{\theequation}
{\Alph{section}.\arabic{equation}}
%
%
%----------------------------------------------------------------------
% Appendix: Proof of the solution in the thermodynamic limit
%----------------------------------------------------------------------
%
\section{Proof of the solution in the thermodynamic limit}
\setcounter{equation}{0}

\hspace{\parindent}In this Appendix we prove the integral representation for 
the one-hole probabilities in the thermodynamic limit
(\ref{ExactIntegralRepresentation}):
\begin{equation}
\label{ExactIntegralCopy}
\Omega(x,y) \;=\; 1-\frac{q^{x+y}}{2 \pi i} \,
\int_{-\infty}^{+\infty} dz\, \Bigl( \frac{1}{z} - \frac{z}{z^2+p^2} \Bigr)
\Bigl( \disp_z^x \disp_{-z}^{y} - \disp_z^{y} \disp_{-z}^{x} \Bigr)\,.
\end{equation}
Instead of deriving this formula from 
the finite-size solutions (\ref{StructF0})-(\ref{StructFR})
by taking $L \rightarrow \infty$ it is much simpler 
to prove that Eq. (\ref{ExactIntegralCopy}) is a solution 
of the one-hole equations (\ref{FullEquations})-(\ref{FullEquationsEnd}). 
We first notice that $\Omega(x,x)=1$
because of the antisymmetry of the integrand. In order to 
verify the bulk equation (\ref{FullEquations}) let us introduce the
notation $g(x,y,z)=q^{x+y}(\disp_z^x \disp_{-z}^{y} - 
\disp_z^{y} \disp_{-z}^{x})$.
Using Eq. (\ref{TheMuFormula}) one can show that
\begin{equation}
q\,g(x-1,y,z)+q^{-1}g(x+1,y,z)+q\,g(x,y-1,z)+q^{-1}g(x,y+1,z) =
2(q+q^{-1}) g(x,y,z)
\end{equation}
This relation implies that Eq. (\ref{ExactIntegralCopy})
satisfies the bulk equation (\ref{FullEquations}). The last step
is to verify the left boundary condition Eq. (\ref{LeftBoundaryCondition}).
Obviously it is equivalent to prove that
\begin{eqnarray}
h(y) &=& q \,\Omega(0,y-1) + q^{-1} \Omega(0,y+1) -
(q+q^{-1}+\inp)\,\Omega(0,y) \\
\nonumber &=& -p \,+\,
\frac{p^2\,q^y}{2\pi} \int \frac{dz}{z}
\Bigl( \frac{\disp_z^y}{z+i\inp} + \frac{\disp_{-z}^y}{z-ip} \Bigr)
\end{eqnarray}
is equal to zero for all $y=1,2,\ldots\infty$. For $y=0$ we get
\begin{equation}
h(0) \;=\; -p \,+\, \frac{p^2}{2} \, \int_{-\infty}^{+\infty}
dz\,\frac{1}{z^2+\inp^2} \;=\; 0
\end{equation}
For $y=1$ one has to solve the integrals
\begin{equation}
h(1) \;=\; -p \,+\, \frac{q\,p^2}{2\pi}\,
\left( \int_{-\infty}^{+\infty} dz\, \frac{\disp_z+\disp_{-z}}{z^2+p^2} -
       \int_{-\infty}^{+\infty} \frac{dz}{z}\, 
             \frac{ip(\disp_z-\disp_{-z})}{z^2+p^2} \right)
\end{equation}
by standard integration techniques in the complex plane. It turns
out that all contributions cancel except at $z=0$ in the second integral
so that $h(1)=0$. Using Eq. (\ref{TheMuFormula}) it is now easy
to derive the recurrence relation
\be
q^{-1}h(y) \,+\, q\,h(y-2) \;=\; (q+q^{-1})\, h(y-1)
\ee
so that $h(y)=0$ for $y=2,3,\ldots\infty$ follows by induction. 
This completes the proof of Eq. (\ref{ExactIntegralCopy}).

\noindent
Let us finally consider the case of infinite input rate where Eq.
(\ref{ExactIntegralCopy}) reduces to
\be
\label{InfiniteSolution}
\Omega(x,y) \;=\; 1 \,-\,
\frac{q^{x+y}}{2 \pi i} \,
\int_{-\infty}^{+\infty} \, \frac{dz}{z} \,
\Bigl(  \disp_{z}^{x} \disp_{-z}^{y} -
	\disp_{z}^{y} \disp_{-z}^{x} \Bigr)\,.
\ee
This expression turns out to be a combination of elliptic integrals. 
This allows us to compute the particle concentration exactly
although the expressions become very complicated as $x$ and $y$
increase. For example the particle concentration in the steady state 
at the first four sites is given by
\begin{eqnarray}
\conc(1) &=& 1 \\
\conc(2) &=& \frac{2}{3 \pi}\Big[\,
	 (q^2+1)(1+6q^2+q^4) \EllipticE \,-\,
	 (q^2+1)(q^2-1)^2 \EllipticK \Big] \,-\, q^2-2q^4 \\
\conc(3) &=& \frac{2}{15 \pi}\Big[\,
	 (q^2+1)(4-15q^2-34q^4-15q^6+4q^8) \EllipticE \\
	 && \hspace{10mm} -\,
	 (q^2+1)(q^2-1)^2(4-15q^2+4q^4) \EllipticK \Big]
	 \,+\, 3q^4-2q^6 \nonumber \\
\conc(4) &=& \frac{4}{105 \pi}\Big[\,
	 (q^2+1)(12-28q^2+45q^4+238q^6+45q^8-28q^{10}+12q^{12})
	 \EllipticE \\ &&  \hspace{12mm} -\, \nonumber
	 (q^2+1)(q^2-1)^2 (12-28q^2+69q^4-28q^6+12q^8)
	 \EllipticK \Big] \,-\, 3q^6-4q^8
\end{eqnarray}
where
\begin{eqnarray}
\EllipticE &=&
\int_0^{\pi/2} d\theta \,
\Big(1-\frac{4}{(q+q^{-1})^2}\sin^2\theta\Big)^{1/2} \\
\EllipticK &=&
\int_0^{\pi/2} d\theta \,
\Big(1-\frac{4}{(q+q^{-1})^2}\sin^2\theta\Big)^{-1/2}
\end{eqnarray}
are elliptic integrals of the first kind.
%
%
%  Appendix: Proof of the scaling limit and function for biased diffusion
%_________________________________________________________________________
%
%
\section{The one point function for biased diffusion to the right in the
continuum limit}
\setcounter{equation}{0}

\hspace{\parindent}In this Appendix we give a proof of equation (\ref{qsf1}). 
We concentrate
on the thermodynamical limit of models in which the particle motion is subject
to a drift pointing away from the source which is situated at the left
boundary ($x=0$) and we are interested in the large $x$ behavior of the 
density.
The starting point is the contour integral (\ref{ContourIntegral}).
In the thermodynamical limit the Green function of the Dirichlet problem
is defined by the $\,i\,=\,j\,=\,0\,$ term of the
multiple sum of (\ref{Green}). We are left with two contributions to the holes 
density function
\be
\Omega^{c}(x,y) \;=\;\Omega^{c}_{\infty}(x,y)\;+\;\Omega^{c}_p(x,y)\,.
\ee
The first one
comes from the integration along the diagonal boundary half-line 
\footnote{We note that the derivatives of the Bessel functions with respect
to their argument are $\,K^{'}_0(u)=\,-K_1(u)\,$ and $\,K^{'}_1(u)=
-\frac12\bigl(K_0(u)+K_2(u)\bigr)$}
$\;(0\,\leq x'\,=\,y'<\,\infty)\;$
\ba
\label{o1}
\Omega^{c}_{\infty}(x,y) &=&\frac{|\pac |}{2 \pi}\,\sum_{\alpha,\beta=\,\pm 1} 
(\alpha y -\beta x)\int_0^{\infty} \frac{K_1(|\pac | r_1)}{r_1}
e^{-\pac [a-(x+y)/2]} \,da
\ea
where $r_1=\sqrt{\Bigl(a-\frac{\alpha x+ \beta y} 2  \Bigr)^2+\Bigl(\frac{
\alpha x- \beta y} 2 \Bigr)^2}$. This is the holes density function in the
case of an infinite input rate.

The second contribution comes from the integration along the left boundary
half-line $\;(0\,\leq y'\,<\,\infty\; , \;x'=0)\,$ and is $\hat p$ dependent 
\be
\label{o3}
\Omega^{c}_p(x,y)\;=\;\frac{|\pac |}{\sqrt 2 \pi}\;\sum_{\alpha =\pm 1} 
\alpha e^{\pac (x+y)/2}\;\int_0^{\infty}\Biggl\{y\frac{K_1\Bigl(
\frac{|\pac | r_2}{\sqrt 2}\Bigr)}{r_2}-x
\frac{K_1\Bigl(\frac{|\pac | r_3}{\sqrt 2}\Bigr)}{r_3}\Biggr\}
\;e^{-\sqrt{\pac ^2/4+\hat{p}}\;\;a} \;\;
da
\ee
where $r_2=\sqrt{\Bigl(a-\alpha x \Bigr)^2+ y^2}$ and 
$r_3=\sqrt{\Bigl(a- \alpha y\Bigr)^2+x^2}$.

The density profile is determined by the holes density function in the limit
$\;y\rightarrow x\;$(see Eq. (\ref{LocalDensity})). 
The behavior of the integrands appearing in (\ref{o1}) and (\ref{o3})
is given by terms of the form $K_1(u)/u$. For $\;u\rightarrow 0\;$ 
the modified Bessel functions
diverge like $\,K_{\nu}(u)\,\sim \,u^{-\nu}$ (for Re$(\nu)>0$).
The only dangerous term is the one containing $r_1$ which vanishes
for $\;y\rightarrow x\;$ and $\;a\rightarrow x\;$ when 
$\alpha=\beta=1$.
The corresponding term in (\ref{o1})
\ba
\label{o0}
\Omega^{c}_0(x,y) &=&\frac{|\pac |(y-x)}{2 \pi}\,
\cdot\int_0^{\infty} \frac{K_1\Biggl(|\pac |\sqrt{\Bigl(a-\frac{x+y} 2 \Bigr)^2+
\frac{(x-y)^2} 4}\; \Biggr)}{\sqrt{\Bigl(a-\frac{x+y} 2 \Bigr )^2+\frac{
(x-y)^2} 4}} e^{-\pac [a-(x+y)/2]} \,da
\ea
determines the asymptotic behavior of the particle concentration, in the 
thermodynamical limit. 
We start with this term. We use the fact that $K_1(\sqrt u)/\sqrt u$ is the 
Laplace transform of $\exp (-1/(4t))\;$ 
(see Eq. (29.3.122) in ref. \cite{Abr}). 
The equation (\ref{o0}) can be rewritten
\be
\label{o02}
\Omega^{c}_0(x,y) \;=\;\pac ^2\frac{y-x}{2 \pi}\,
\cdot\int_0^{\infty}
\exp \Biggl( -t \pac ^2 \Bigl(\frac {x-y}{2}\Bigr)^2 \Biggr) \;dt\;
\int_{-\frac{x+y}{2}}^{\infty}
\exp \Biggl( -t \Bigl(\pac  a +\frac {1}{2t}\Bigr)^2 \Biggr) \;da\;
\ee
After integrating over the variable $a$ we get
\be
\label{o03}
\Omega^{c}_0(x,y) \;=\;1\,-\,\pac \frac{y-x}{2 \sqrt{\pi}}\,
\cdot\int_0^{\infty}
\exp \Biggl( -t^2 \pac ^2 \Bigl(\frac {x-y}{2}\Bigr)^2 \Biggr) \;
\mbox{erfc}\Bigl(t \pac  \frac{x+y}{2}-\frac{1}{2t}\Bigr) \;dt
\ee
Here $\mbox{erfc}$ stands for the complementary error function. From 
(\ref{LocalDensity}) we get that the contribution of $\Omega^{c}_0(x,y)$
to the particle density is
\be
\label{r01}
\density_0(x) \;=\;\frac{\pac }{2 \sqrt{\pi}}\,
\cdot\int_0^{\infty}
\;\mbox{erfc}\Bigl(t \pac  x-\frac{1}{2t}\Bigr)
\;dt
\ee
With the change of variable 
\be
t\;=\;\frac{\omega+\sqrt{\omega ^2+2 \pac  x}}{2 \pac  x}
\ee
we get
\be
\label{r02}
\density_0(x) \;=\;\frac{1}{2 \sqrt{\pi} x}\,\Biggl(
\int_0^{\infty}
\Bigl(1-\frac{\omega}{\sqrt{\omega^2+2\pac  x}}\Bigr) \;d\omega\;\;\;
+\;\;\;\int_0^{\infty}\,\frac{\omega}{
\sqrt{\omega^2+2\pac  x}}\mbox{erfc}(\omega)\;d\omega\;\Biggr)
\ee
After integrating by parts one gets
\be
\density_0(x)\;=\;\frac 1 {\pi x} \,\int_0^{\infty}\,
\sqrt{\omega^2 + 2\pac  x}\,e^{-\omega^2}\,d\omega
\ee
Expanding in powers of $\omega^2/x$ we finally obtain:  
\be
\density_0(x)\;=\;\sqrt{\frac \pac  {2\pi x}} \,+\,\frac 1 {8\sqrt{2\pi \pac }}
\frac 1 {x^{3/2}}\;+\;O(x^{-5/2})
\label{densv1}
\ee
The rest of the holes density function
\be
\label{rest}
\Omega^{c}_{rest}(x,y)\;=\;\Omega^{c}(x,y)-\Omega^{c}_0(x,y)
\ee
gives also a contribution to the particle density
\be
\density_{rest}(x) \;=\;- \left. \frac{\partial}{\partial y}\Omega^{c}_{rest}
(x,y)\right|_{y=x}
\label{B6}
\ee
In all the integrals contributing to $\Omega_{rest}^c$, the arguments of the 
Bessel functions ($r_i,\,i=1,2,3\,$) are greater then $x$ (or $y$).
We can thus use the asymptotic expansions of $\,K_{\nu}(u)\;$ and replace
these functions with $\,\exp(-u)\cdot\sqrt{\pi/(2u)}\;$.
One obtains an integral expression of $\,\density_{rest}(x)$. Due to the 
exponential falloff of the integrands appearing in (\ref{B6}),
one can expand them in powers of $a/x$ (where $a$ is the integration variable).
After some computations one gets that for $x\,\rightarrow\,\infty\;$
$\;\density_{rest}(x)\;$ decays like $x^{-3/2}$. 

\noindent
Summing up we get the following formula for the asymptotic behavior of the
density profile
\be
\density (x)\;=\;\sqrt{\frac \pac  {2\pi x}} \,+\,\frac 1 {4\sqrt{2\pi \pac }}
\Biggl[ \frac {11} 2 \,-\,\frac{\pac ^4}{\hat{p}^2} \Biggr]
\frac 1 {x^{3/2}}\;+\;O(x^{-5/2})
\label{densv}
\ee
We notice that the leading term is independent of the input rate $\hat{p}$
but the next to leading term is $\hat{p}$ dependent.
%
%  Appendix: Technical details about the Monte Carlo simulations
%_________________________________________________________________________
%
%
\section{Details on the Monte Carlo simulations}
\setcounter{equation}{0}

\hspace{\parindent}In this Appendix we explain how we have done the simulations
of the coagulation-diffusion model with particle input at one boundary 
($p_R=0$). We can simplify the notation and use the symbol $\;p\;$ instead of 
$p_L$ for the input rate.

The simplest way to simulate reaction-diffusion models is to use a Monte Carlo
algorithm with random sequential updates. However, this algorithm is
not very efficient for the present problem since particle densities are very 
low and therefore
most of the updates take place at empty sites. This is why we used
a different method in which the positions of the particles are stored
rather than the occupation numbers of the sites (for details see \cite{Krebs2} 
and references therein). This 'direct' method is much faster than the first 
one. It is defined as follows. At  the beginning the lattice is empty.
As long as the total number of particles in the system is $0$
the following steps are repeated:

\begin{itemize}
\item choose a randomly a number $a$, between $0$ and $1$
\item occupy site 1 with a particle if $a \leq p\triangle t$
\item leave the site unoccupied if $a > p\triangle t$
\item increment the time $t \rightarrow t + \triangle t$
\end{itemize}

\noindent
The parameter $\triangle t$ is the time discretization (see Ref. \cite{Krebs2}).
After the first particle entered the system, the 'direct' Monte-Carlo
algorithm is started:
\begin{itemize}
\item choose a {\em particle} at random and one of its neighboring sites
\item update the configuration of the chosen pair with help of
      a random number and by considering the bulk reaction rates
\item increment the time $t \rightarrow t + \frac{\triangle t}{N}$ where $N$ is
      the current total number of particles in the system
\item if site $1$ is empty try to occupy it by comparing a random
      number  with $p\frac{\triangle t}{N}$. If $\,p\,=\,\infty\,$ one 
	keeps the site $1$ occupied at all time-steps.
\end{itemize}

In order to test the accuracy of the ``direct'' Monte Carlo method, we simulated
some systems for which analytical data is available. The agreement
is very good. For $q=1$ we used a lattice of length $L=200$ and took
$p=1$ and $p=0.01$. Only for the first $10-15$ sites the two sets of
values for the density profile are slightly different (the relative 
difference is  of less than $10\%\,$). For the other sites the difference
between the two measurements of the density profile is zero
within numerical errors.
We also compared data obtained for systems characterized by $q=\sqrt{2.5}$,
$\,L\,=\,20\,$ sites and $p=1$ and $p=0.1$. Although the lattice length
is small, the two sets of measurements of the density profiles coincide for
all sites within numerical errors.

The quality of the Monte Carlo simulations is higher in the case where
the particle diffusion is biased to the right in comparison with the 
$a_R\;\leq\;a_L\;$ case. This has two reasons. On the 
one hand in the $a_R\;>a_L$ case the total concentration
of particles in the stationary state is larger
as compared to the symmetric case. 
It is more likely to reproduce through simulations a distribution with a
higher total number of particles.
On the other hand the relaxation of these system to the 
stationary state occurs much faster than in the $a_R\;=\;a_L\;$ case since the
time operator has massive excitations for $a_R\;\neq\;a_L\;$. 
Therefore less CPU time per run is necessary and thus the numerical errors
of the measurements are smaller.

It is a well known fact that the quality of the Monte Carlo determinations is 
limited by the accuracy of the random number generator. If the number of 
steps requiring random numbers is too high, at some point the generator
produces correlated numbers. This limitation poses some problems in the case of
symmetric diffusion, for large lattices. This explains the unphysical
oscillations of the data corresponding to $L=1000$ in Figures
\ref{f2} and \ref{f3}. The choice of a better random number generator 
implies the increase of the CPU time needed to perform the simulations.

Since we are interested in the stationary properties of the system
we stopped each simulation run at a value of $t\,=\,t_{max}\,$ such that
at least in the time interval $\,[t_{max}/3\, ,\,t_{max}]\,$ the average total
number of particles is fluctuating around a constant value. 
We used a double averaging technique. We took $100$ equidistant time points
between$\,[0.9\cdot t_{max}\, ,\,t_{max}]\,$ and measured our observables in 
each of them. For each Monte Carlo run of the program we got a preliminary value
by averaging over this $100$ determinations. Afterwards we averaged these
preliminary values over all MC runs. The number of runs performed for each
system was between $4$ and $50$ thousand, depending on the lattice length.
Due to CPU time limitations, the number of runs performed decreases
with the lattice length.

For the data presented in Figs. \ref{f2} and \ref{f3} we used coarse-graining 
for obvious reasons, this is reflected in the horizontal error bars.
%
%
%----------------------------------------------------------------------
% References
%----------------------------------------------------------------------
%
%

\listoftables

\newcounter{fig_count}
\vspace{1cm}
\noindent {\Large \bf List of Figures}

\vspace{1cm}

\begin{list}
      {Fig. \arabic{fig_count}:}
      {\usecounter{fig_count}  \setlength{\leftmargin}{1.5cm}
               \setlength{\labelsep}{2mm}
               \setlength{\labelwidth}{1.3cm}}
\item
\label{f0}
Bulk and boundary processes in the coagulation-diffusion model

\item 
\label{f1}
The function $F(z)$ defined by Eq. (\ref{rsf}) 
(solid curve) compared to lattice calculations for $L=2000$ ($p=1.0$) and
$L=800$ ($p=\infty$).
	
\item 
\label{f2}
The $L$ dependence of the $K(z,L$) function define 
by Eq. (\ref{kzl}) for $p=1.0,\, \pb\,=\,0.5,\,\pa = \pc =0\;$. If
the scaling function is universal, $K(z,L)$ should vanish in the 
thermodynamical limit. (Monte Carlo simulations)

%	Figure \label{f3} The $K(z,L$ function for $L=1000$ and different
%\addcontentsline{lof}{Figure}{The $K(z,L$ function for $L=1000$ and different
\item 
\label{f3}
The $K(z,L)$ function for $L=1000,\;\;\pa=0$ and different
input and coagulation rates. (Monte Carlo simulations)

\end{list}
%
%----------------------------------------------------------------------
% FIGURES
%----------------------------------------------------------------------
%
% FIGURE 2
%=========
\newpage
\begin{figure}[h]
\epsfxsize=140mm
\epsffile{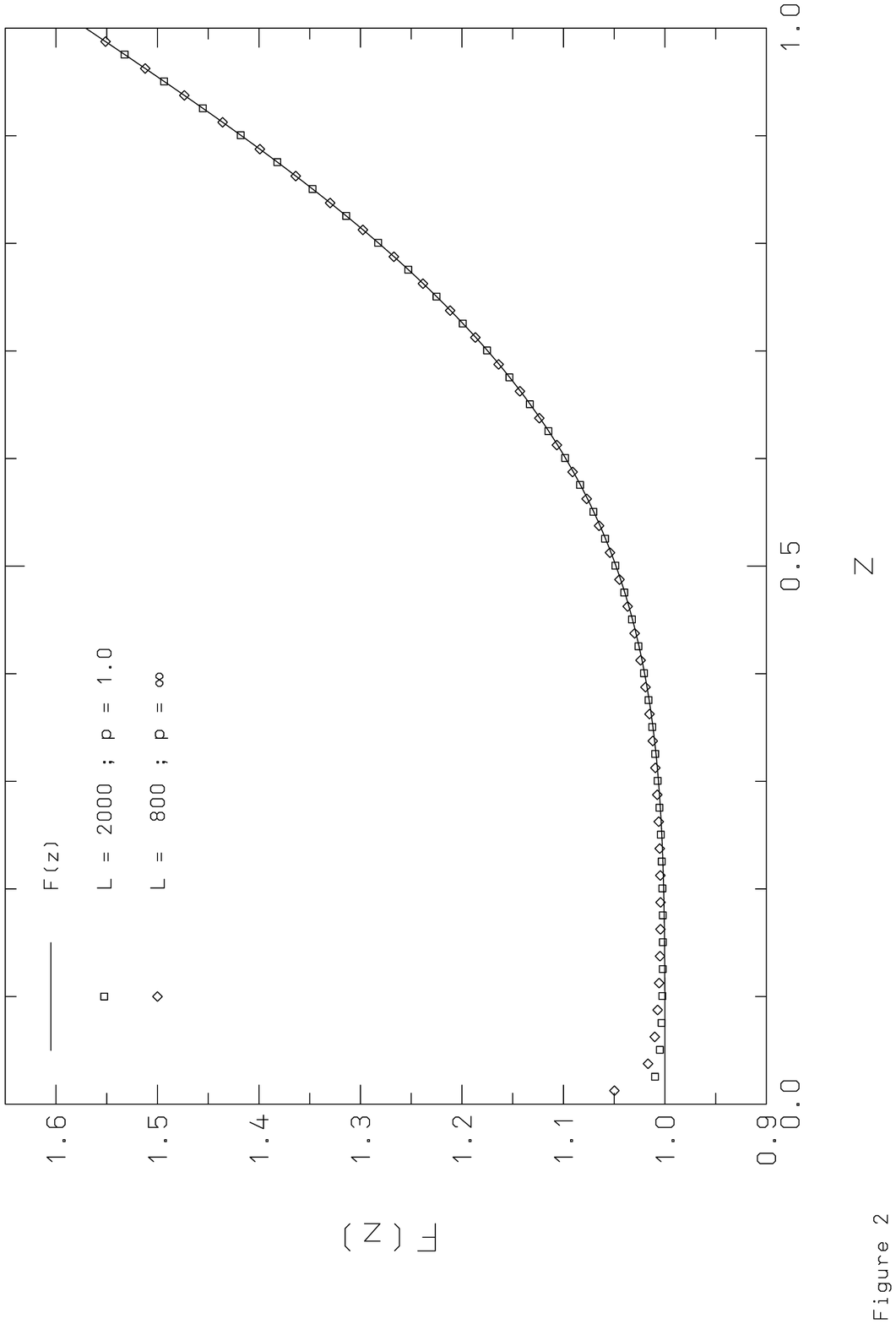}
\end{figure}
%
% FIGURE 3
%=========
\newpage
\begin{figure}[h]
\epsfxsize=140mm
\epsffile{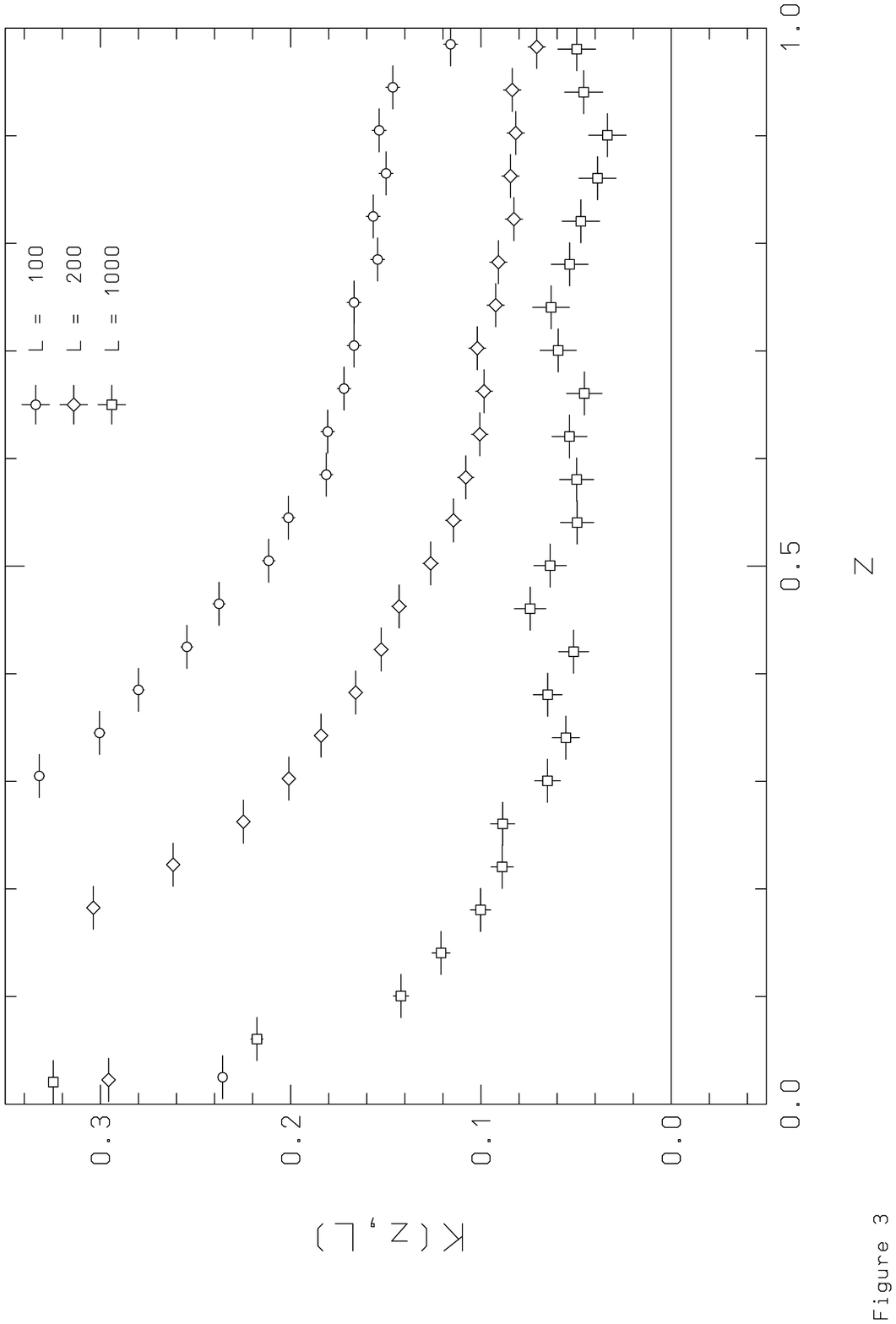}
\end{figure}
%
% FIGURE 4
%=========
\newpage
\begin{figure}[h]
\epsfxsize=140mm
\epsffile{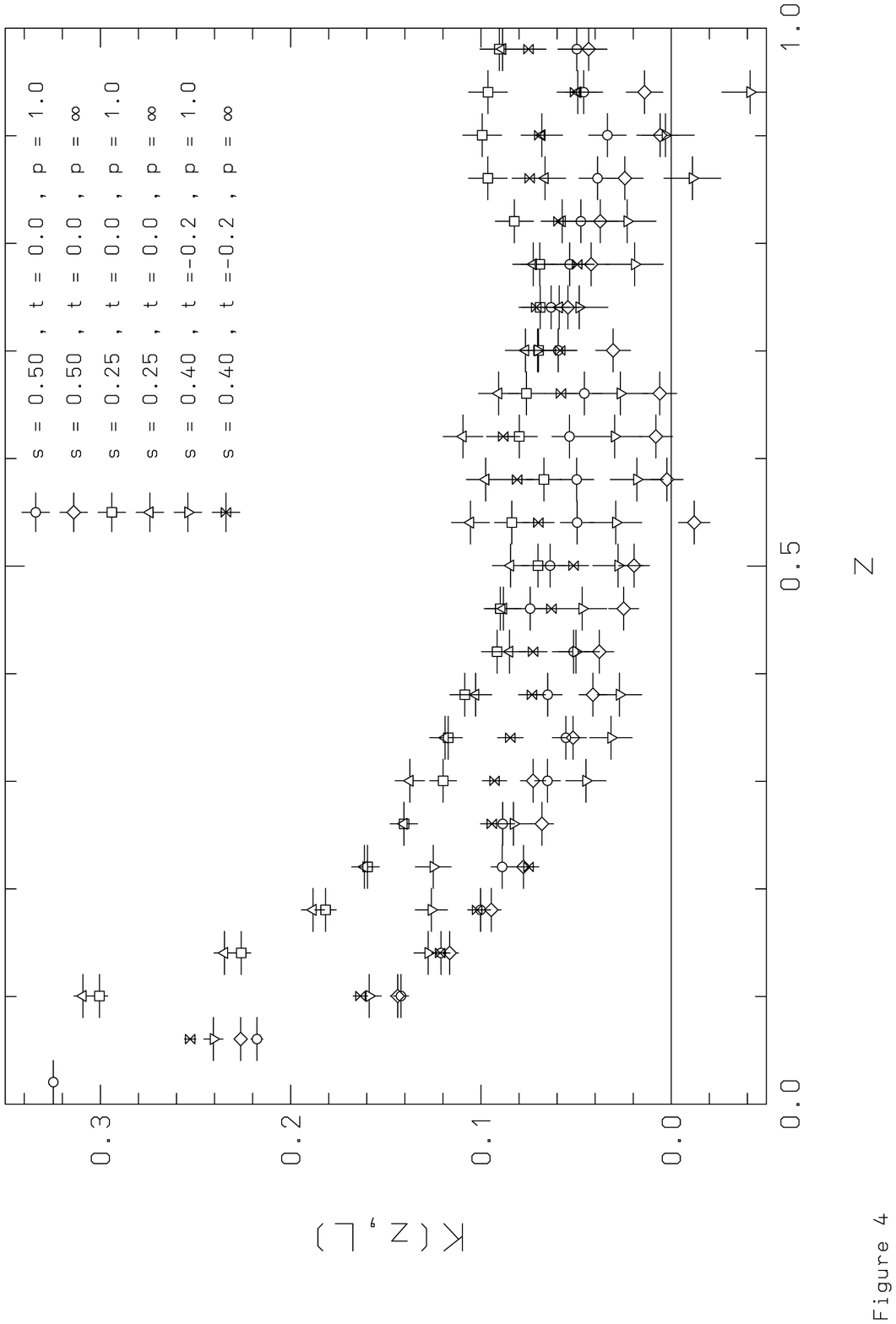}
\end{figure}
\end{document}